\documentclass{ws-ijgmmp}
\usepackage[utf8]{inputenc}
\usepackage[T1]{fontenc}
\usepackage[colorlinks,hyperindex,plainpages=false]{hyperref}
\usepackage{amsmath}
\usepackage{amssymb}
\usepackage{amsfonts}
\usepackage{accents}

\DeclareMathAlphabet{\mathds}{U}{BOONDOX-ds}{m}{n}
\DeclareMathOperator{\sinc}{sinc}
\DeclareMathOperator{\arcsinh}{arcsinh}

\newcommand{\dd}{\mathrm{d}}
\newcommand{\DD}{\mathrm{D}}
\newcommand{\intprod}{\mathrel{\reflectbox{\rotatebox[origin=c]{180}{$\neg$}}}}
\newcommand{\lc}[1]{\accentset{\circ}{#1}}
\DeclareMathOperator{\Vect}{Vect}

\begin{document}

\markboth{Manuel Hohmann}
{Complete classification of cosmological teleparallel geometries}

%
\catchline{}{}{}{}{}
%

\title{Complete classification of cosmological teleparallel geometries}

\author{Manuel Hohmann}

\address{Laboratory of Theoretical Physics, Institute of Physics, University of Tartu, W. Ostwaldi 1, 50411 Tartu, Estonia\\
\email{manuel.hohmann@ut.ee}}

\maketitle

\begin{history}
\received{(Day Month Year)}
\revised{(Day Month Year)}
\end{history}

\begin{abstract}
We consider the notion of cosmological symmetry, i.e., spatial homogeneity and isotropy, in the field of teleparallel gravity and geometry, and provide a complete classification of all homogeneous and isotropic teleparallel geometries. We explicitly construct these geometries by independently employing three different methods, and prove that all of them lead to the same class of geometries. Further, we derive their properties, such as the torsion tensor and its irreducible decomposition, as well as the transformation behavior under change of the time coordinate, and derive the most general cosmological field equations for a number of teleparallel gravity theories. In addition to homogeneity and isotropy, we extend the notion of cosmological symmetry to also include spatial reflections, and find that this further restricts the possible teleparallel geometries. This work answers an important question in teleparallel cosmology, in which so far only particular examples of cosmologically symmetric solutions had been known, but it was unknown whether further solutions can be constructed.
\end{abstract}


\section{Introduction}\label{sec:intro}
Some of the most prominent open questions in modern physics arise from observations in cosmology, which hint towards phases of accelerating expansion in both the early and late history of the universe. An explanation of these observations requires either the introduction of a new matter type, known as dark energy, whose properties significantly differ from any matter thus far observed in particle physics, or a modification of our description of gravity by general relativity. The latter is also suggested by the tensions between general relativity and quantum physics, which have so far obstructed the construction of a complete and conclusive theory of quantum gravity.

While most approaches to modify general relativity are based on its standard formulation in terms of the curvature of the Levi-Civita connection of the spacetime metric, there exist also approaches based on other geometries, known as the metric teleparallel and symmetric teleparallel formulations~\cite{BeltranJimenez:2019tjy}, in which gravity is mediated by the torsion or nonmetricity of a flat connection instead. In this article we will focus on the metric teleparallel approach (and omit the word ``metric'' is this context for brevity) and consider geometries based on a metric and a metric-compatible, flat, affine connection, which is characterized by its torsion~\cite{Aldrovandi:2013wha}. Another, more common description of such geometries is the Lorentz covariant formulation, which makes use of a tetrad and a flat, antisymmetric spin connection~\cite{Krssak:2015oua,Krssak:2018ywd}.

In order to study the cosmological dynamics of any theory of gravity, one must apply its field equations to a spacetime geometry which obeys the cosmological symmetry, being spatially homogeneous and isotropic. In curvature-based theories, which employ a pseudo-Riemannian geometry fully defined by the metric tensor, this is easily implemented by making use of the Friedmann-Lemaitre-Robertson-Walker (FLRW) metric, which is the most general cosmologically symmetric metric. In teleparallel gravity, however, this is not sufficient in general, since the gravitational dynamics are also influenced by further degrees of freedom beyond the metric ones, which are encoded in the tetrad and the spin connection, and enter the field equations through the torsion of the teleparallel connection. Failure to consider a ``proper'' teleparallel geometry may result in the field equations to reduce to that of general relativity, excluding any modifications~\cite{Tamanini:2012hg}. Finding ``proper'' teleparallel geometries which allow the study of different types of solutions of modified teleparallel gravity theories and the resulting deviations from general relativity has therefore become an important task in the field of teleparallel gravity.

An important step for the construction of suitable geometries to study teleparallel gravity has been achieved by realizing that this task is simplified if one considers geometries which obey an extended notion of symmetry, which takes into account not only the metric degrees of freedom, but also the teleparallel connection, or equivalently, the tetrad and spin connection~\cite{Hohmann:2019nat}. For the particular case of cosmological symmetry, it has been shown that any teleparallel geometry which satisfies these symmetry conditions automatically solves the antisymmetric part of the field equations of any teleparallel gravity theory, hence significantly simplifying the construction of explicit cosmological solutions. Various examples of such cosmologically symmetric teleparallel geometries have been found, and their properties as solutions to teleparallel gravity theories have been studied~\cite{Tamanini:2012hg,Hohmann:2019nat,Ferraro:2011us,Capozziello:2018hly,Hohmann:2018rwf}. However, it has so far been an open question whether one may find any further teleparallel geometries which obey the cosmological symmetry.

In this article we answer this question by explicitly constructing the most general class of cosmologically symmetric teleparallel geometries. For this purpose we employ three different, independent methods: besides the symmetry condition of the tetrad and spin connection~\cite{Hohmann:2019nat} we also use the symmetry of a metric-affine geometry~\cite{Hohmann:2019fvf}, and a novel method based on the irreducible decomposition of the torsion tensor~\cite{Hehl:1994ue}. One purpose of this threefold approach is illustrative, by demonstrating how to use these different methods, while at the same time we prove their mutual consistency by showing that they lead to the same result. Further, we show how different cosmologically symmetric geometries can be related to each other by coordinate transformations, and derive the resulting cosmological field equations for a number of teleparallel gravity theories.

The article is structured as follows. In section~\ref{sec:pre}, we review the relevant physical and mathematical notions we use in our derivation. The main part of this article, the construction of the most general cosmologically symmetric teleparallel geometries, is the presented in section~\ref{sec:construction}. The following sections discuss properties of the obtained geometries. In section~\ref{sec:cotrans} we show how different teleparallel geometries in the cosmologically symmetric class are related by coordinate transformations. Their application to teleparallel gravity theories and derivation of the resulting cosmological dynamics is shown in section~\ref{sec:dynamics}. We end with a conclusion in section~\ref{sec:conclusion}.

\section{Preliminaries}\label{sec:pre}
Before we come to the main part of this article, we briefly review a few physical and mathematical notions we will be using, and establish notational conventions. In section~\ref{ssec:telegeo}, we give an overview over the notion of teleparallel geometry both in its metric-affine and tetrad / spin connection representations. The notion of spacetime symmetries for teleparallel geometries is reviewed in section~\ref{ssec:stsym}. We then focus on cosmological symmetry in section~\ref{ssec:cosmosym}, where we provide our conventions for the coordinates and symmetry generators we will use. Conventions for the Lorentz group are established in section~\ref{ssec:loralgeb}. We conclude with a remark on complexified tetrads in section~\ref{ssec:complex}.

\subsection{Teleparallel geometry}\label{ssec:telegeo}
We begin with a brief review of the different possibilities to describe a teleparallel geometry which we will use in this article. The most common description of teleparallel geometry on a four-dimensional spacetime manifold \(M\) is given by a tetrad (or coframe) \(\theta^A = \theta^A{}_{\mu}\dd x^{\mu}\) and a spin connection \(\omega^A{}_B = \omega^A{}_{B\mu}\dd x^{\mu}\), where we use capital Latin letters \(A, B = 0, \ldots 3\) to denote Lorentz indices, while lowercase Greek indices \(\mu, \nu = 0, \ldots, 3\) will denote spacetime indices. We demand that at each point \(x \in M\) the tetrad is invertible, and we denote its inverse (the frame) by \(e_A = e_A{}^{\mu}\partial_{\mu}\), satisfying \(\theta^A{}_{\mu}e_A{}^{\nu} = \delta_{\mu}^{\nu}\) and \(\theta^A{}_{\mu}e_B{}^{\mu} = \delta^A_B\). Further, we impose two conditions on the spin connection. The first is the metricity condition, which can most easily written in terms of vanishing nonmetricity
\begin{equation}\label{eq:nmform}
Q_{AB} = \DD\eta_{AB} = \eta_{CB}\omega^C{}_A + \eta_{AC}\omega^C{}_{B} \equiv 0\,.
\end{equation}
Here we have introduced the Minkowski metric \(\eta = \mathrm{diag}(-1, 1, 1, 1)\) and the exterior covariant derivative
\begin{equation}
\begin{split}
\DD W^{A_1 \cdots A_r}{}_{B_1 \cdots B_s} &= \dd W^{A_1 \cdots A_r}{}_{B_1 \cdots B_s}\\
&\phantom{=}+ \omega^{A_1}{}_C \wedge W^{CA_2 \cdots A_r}{}_{B_1 \cdots B_s} + \ldots + \omega^{A_r}{}_C \wedge W^{A_1 \cdots A_{r - 1}C}{}_{B_1 \cdots B_s}\\
&\phantom{=}- \omega^C{}_{B_1} \wedge W^{A_1 \cdots A_r}{}_{CB_2 \cdots B_s} + \ldots - \omega^C{}_{B_s} \wedge W^{A_1 \cdots A_r}{}_{B_1 \cdots B_{s - 1}C}\,.
\end{split}
\end{equation}
The second condition we impose on the spin connection is that of vanishing curvature
\begin{equation}\label{eq:curvform}
R^A{}_B = \dd\omega^A{}_B + \omega^A{}_C \wedge \omega^C{}_B \equiv 0\,.
\end{equation}
Finally, we define the (in general non-vanishing) torsion as
\begin{equation}\label{eq:torsform}
T^A = \DD\theta^A = \dd\theta^A + \omega^A{}_B \wedge \theta^B\,.
\end{equation}
It is also possible to describe the teleparallel geometry in terms of a metric \(g_{\mu\nu}\) and an affine connection \(\nabla\) with coefficients \(\Gamma^{\mu}{}_{\nu\rho}\). In terms of the previously defined quantities they are given by the relations
\begin{equation}\label{eq:metric}
g_{\mu\nu} = \eta_{AB}\theta^A{}_{\mu}\theta^B{}_{\nu}
\end{equation}
and
\begin{equation}\label{eq:affconn}
\Gamma^{\mu}{}_{\nu\rho} = e_A{}^{\mu}\left(\partial_{\rho}\theta^A{}_{\nu} + \omega^A{}_{B\rho}\theta^B{}_{\nu}\right)\,.
\end{equation}
In these variables, the condition of vanishing nonmetricity reads
\begin{equation}\label{eq:affnonmet}
Q_{\rho\mu\nu} = \nabla_{\rho}g_{\mu\nu} \equiv 0\,,
\end{equation}
while vanishing curvature is expressed as
\begin{equation}\label{eq:affcurv}
R^{\rho}{}_{\sigma\mu\nu} = \partial_{\mu}\Gamma^{\rho}{}_{\sigma\nu} - \partial_{\nu}\Gamma^{\rho}{}_{\sigma\mu} + \Gamma^{\rho}{}_{\tau\mu}\Gamma^{\tau}{}_{\sigma\nu} - \Gamma^{\rho}{}_{\tau\nu}\Gamma^{\tau}{}_{\sigma\mu} \equiv 0\,.
\end{equation}
Also the torsion is then more conveniently expressed purely in tensor components in the form
\begin{equation}\label{eq:torsion}
T^{\rho}{}_{\mu\nu} = \Gamma^{\rho}{}_{\nu\mu} - \Gamma^{\rho}{}_{\mu\nu}\,.
\end{equation}
It is related to the torsion two-form~\eqref{eq:torsform} via
\begin{equation}
T^A = \frac{1}{2}T^A{}_{\mu\nu}\dd x^{\mu} \wedge \dd x^{\nu} = \frac{1}{2}\theta^A{}_{\rho}T^{\rho}{}_{\mu\nu}\dd x^{\mu} \wedge \dd x^{\nu}\,.
\end{equation}
It is further helpful to define the contortion tensor
\begin{equation}\label{eq:contor}
K^{\mu}{}_{\nu\rho} = \frac{1}{2}\left(T_{\nu}{}^{\mu}{}_{\rho} + T_{\rho}{}^{\mu}{}_{\nu} - T^{\mu}{}_{\nu\rho}\right)\,,
\end{equation}
through which the coefficients \(\Gamma^{\mu}{}_{\nu\rho}\) of the teleparallel connection are given as
\begin{equation}\label{eq:affconndec}
\Gamma^{\mu}{}_{\nu\rho} = \lc{\Gamma}^{\mu}{}_{\nu\rho} + K^{\mu}{}_{\nu\rho}\,,
\end{equation}
where \(\lc{\Gamma}^{\mu}{}_{\nu\rho}\) are the coefficients of the Levi-Civita connection of the metric \(g_{\mu\nu}\).

\subsection{Spacetime symmetries}\label{ssec:stsym}
We briefly review the notion of spacetime symmetries of metric-affine geometries, and hence in particular teleparallel geometries, which we use in this article; see~\cite{Hohmann:2019nat,Hohmann:2015pva} for a detailed discussion. Here and in the remainder of this article we will assume the (left) action \(\varphi: G \times M \to M\) of a Lie group \(G\) on the spacetime manifold. For \(u \in G\), we denote the induced diffeomorphism by \(\varphi_u: M \to M\). Under these diffeomorphisms, the metric and connection coefficients of a metric-affine geometry transform via their pullbacks as
\begin{equation}
(\varphi_u^*g)_{\mu\nu}(x) = g_{\tau\omega}(x')\frac{\partial x'^{\tau}}{\partial x^{\mu}}\frac{\partial x'^{\omega}}{\partial x^{\nu}}
\end{equation}
and
\begin{equation}
(\varphi_u^*\Gamma)^{\mu}{}_{\nu\rho}(x) = \Gamma^{\sigma}{}_{\tau\omega}(x')\frac{\partial x^{\mu}}{\partial x'^{\sigma}}\frac{\partial x'^{\tau}}{\partial x^{\nu}}\frac{\partial x'^{\omega}}{\partial x^{\rho}} + \frac{\partial x^{\mu}}{\partial x'^{\sigma}}\frac{\partial^2x'^{\sigma}}{\partial x^{\nu}\partial x^{\rho}}\,,
\end{equation}
while the tetrad and spin connection transform as one-forms, hence
\begin{equation}
(\varphi_u^*\theta)^A{}_{\mu}(x) = \theta^A{}_{\nu}(x')\frac{\partial x'^{\nu}}{\partial x^{\mu}}
\end{equation}
and
\begin{equation}
(\varphi_u^*\omega)^A{}_{B\mu}(x) = \omega^A{}_{B\nu}(x')\frac{\partial x'^{\nu}}{\partial x^{\mu}}\,.
\end{equation}
We say that a metric-affine geometry is \emph{symmetric} under the group action \(\varphi\) if and only if for every \(u \in G\) the metric and connection are invariant,
\begin{equation}
(\varphi_u^*g)_{\mu\nu} = g_{\mu\nu}\,, \quad
(\varphi_u^*\Gamma)^{\mu}{}_{\nu\rho} = \Gamma^{\mu}{}_{\nu\rho}\,.
\end{equation}
For the tetrad and spin connection it then follows that the induced metric-affine geometry is symmetric if and only if for every \(u \in G\) there exists a local Lorentz transformation \(\boldsymbol{\Lambda}_u: M \to \mathrm{SO}(1,3)\) such that
\begin{equation}\label{eq:tetfinsym}
(\varphi_u^*\theta)^A{}_{\mu}(x) = (\boldsymbol{\Lambda}_u^{-1})^A{}_B(x)\theta^B{}_{\mu}(x)
\end{equation}
and
\begin{equation}\label{eq:spcfinsym}
(\varphi_u^*\omega)^A{}_{B\mu}(x) = (\boldsymbol{\Lambda}_u^{-1})^A{}_C(x)\left[\boldsymbol{\Lambda}_u^D{}_B(x)\omega^C{}_{D\mu}(x) + \partial_{\mu}\boldsymbol{\Lambda}_u^C{}_B(x)\right]\,.
\end{equation}
Consistency between successively applied diffeomorphisms for \(u, v\) leads to
\begin{equation}
(\boldsymbol{\Lambda}_{uv}^{-1})^A{}_B\theta^B{}_{\mu} = (\varphi_{uv}^*\theta)^A{}_{\mu} = (\varphi_v^*\varphi_u^*\theta)^A{}_{\mu} = (\boldsymbol{\Lambda}_v^{-1})^A{}_B(\boldsymbol{\Lambda}_u^{-1})^B{}_C\theta^C{}_{\mu}\,,
\end{equation}
and so \(\boldsymbol{\Lambda}: G \times M \to \mathrm{SO}(1,3)\) must be a local group homomorphism, \(\boldsymbol{\Lambda}_{uv} = \boldsymbol{\Lambda}_u\boldsymbol{\Lambda}_v\).

For practical purposes it is often more convenient to work with infinitesimal symmetries instead of finite diffeomorphisms. For the Lie group action discussed above, they are described by the fundamental vector fields \(X: \mathfrak{g} \to \Vect(M)\), where \(\mathfrak{g}\) denotes the Lie algebra of \(G\), while \(\Vect(M)\) denotes the space of vector fields on \(M\). Under an infinitesimal transformation \(X_{\xi} \in \Vect(M)\) with \(\xi \in \mathfrak{g}\) the metric and affine connection transform via their Lie derivatives as
\begin{equation}
(\mathcal{L}_{X_{\xi}}g)_{\mu\nu} = X_{\xi}^{\rho}\partial_{\rho}g_{\mu\nu} + \partial_{\mu}X_{\xi}^{\rho}g_{\rho\nu} + \partial_{\nu}X_{\xi}^{\rho}g_{\mu\rho}
\end{equation}
and
\begin{equation}
\begin{split}
(\mathcal{L}_{X_{\xi}}\Gamma)^{\mu}{}_{\nu\rho} &= X_{\xi}^{\sigma}\partial_{\sigma}\Gamma^{\mu}{}_{\nu\rho} - \partial_{\sigma}X_{\xi}^{\mu}\Gamma^{\sigma}{}_{\nu\rho} + \partial_{\nu}X_{\xi}^{\sigma}\Gamma^{\mu}{}_{\sigma\rho} + \partial_{\rho}X_{\xi}^{\sigma}\Gamma^{\mu}{}_{\nu\sigma} + \partial_{\nu}\partial_{\rho}X_{\xi}^{\mu}\\
&= \nabla_{\rho}\nabla_{\nu}X_{\xi}^{\mu} - X_{\xi}^{\sigma}R^{\mu}{}_{\nu\rho\sigma} - \nabla_{\rho}(X_{\xi}^{\sigma}T^{\mu}{}_{\nu\sigma})\,,
\end{split}
\end{equation}
where we included the last line as a reminder that this expression is a tensor. For the tetrad and spin connection, the corresponding transformations read
\begin{equation}
(\mathcal{L}_{X_{\xi}}\theta)^A{}_{\mu} = X_{\xi}^{\nu}\partial_{\nu}\theta^A{}_{\mu} + \partial_{\mu}X_{\xi}^{\nu}\theta^A{}_{\nu}
\end{equation}
and
\begin{equation}
(\mathcal{L}_{X_{\xi}}\omega)^A{}_{B\mu} = X_{\xi}^{\nu}\partial_{\nu}\omega^A{}_{B\mu} + \partial_{\mu}X_{\xi}^{\nu}\omega^A{}_{B\nu}\,.
\end{equation}
Following the same line of thought as for the finite case, we say that the fundamental vector fields generate an infinitesimal symmetry of the metric-affine geometry if and only if for all \(\xi \in \mathfrak{g}\) the Lie derivatives of the metric and affine connection vanish,
\begin{equation}
(\mathcal{L}_{X_{\xi}}g)_{\mu\nu} = 0\,, \quad
(\mathcal{L}_{X_{\xi}}\Gamma)^{\mu}{}_{\nu\rho} = 0\,.
\end{equation}
This is the case if and only if for every \(\xi \in \mathfrak{g}\) there exists an infinitesimal local Lorentz transformation \(\boldsymbol{\lambda}_{\xi}: M \to \mathfrak{so}(1,3)\) such that
\begin{equation}\label{eq:tetinfsym}
(\mathcal{L}_{X_{\xi}}\theta)^A{}_{\mu} = -\boldsymbol{\lambda}_{\xi}^A{}_B\theta^B{}_{\mu}
\end{equation}
and
\begin{equation}\label{eq:spcinfsym}
(\mathcal{L}_{X_{\xi}}\omega)^A{}_{B\mu} = \partial_{\mu}\boldsymbol{\lambda}_{\xi}^A{}_B + \omega^A{}_{C\mu}\boldsymbol{\lambda}_{\xi}^C{}_B - \omega^C{}_{B\mu}\boldsymbol{\lambda}_{\xi}^A{}_C\,.
\end{equation}
Also in this case consistency between successive transformations requires that the map \(\boldsymbol{\lambda}: \mathfrak{g} \times M \to \mathfrak{so}(1,3)\) is not arbitrary, but must be a local Lie algebra homomorphism, \(\boldsymbol{\lambda}_{[\xi, \zeta]} = [\boldsymbol{\lambda}_{\xi}, \boldsymbol{\lambda}_{\zeta}]\).

Finally, we remark that in the Weitzenböck gauge \(\omega^A{}_{B\mu} \equiv 0\) the symmetry conditions~\eqref{eq:spcfinsym} and~\eqref{eq:spcinfsym} become particularly simple, and read
\begin{equation}
\partial_{\mu}\boldsymbol{\Lambda}_u^A{}_B = 0\,, \quad
\partial_{\mu}\boldsymbol{\lambda}_{\xi}^A{}_B = 0\,,
\end{equation}
so that \(\boldsymbol{\Lambda}\) and \(\boldsymbol{\lambda}\) reduce to a global Lie group and Lie algebra homomorphisms, respectively. It will be a crucial part of our derivation presented in this article to classify all such homomorphisms for the case of the cosmological symmetry groups.

\subsection{Cosmological symmetry}\label{ssec:cosmosym}
In the remainder of this article, we will consider as symmetry groups \(G\) the groups representing cosmological symmetry, i.e., homogeneity and isotropy. Their infinitesimal action on the spacetime manifold is described in spherical coordinates \((t,r,\vartheta,\varpi)\) in terms of the generating vector fields of rotations
\begin{subequations}\label{eq:genrot}
\begin{align}
R_1 &= \sin\varphi\partial_{\vartheta} + \frac{\cos\varphi}{\tan\vartheta}\partial_{\varphi}\,,\\
R_2 &= -\cos\varphi\partial_{\vartheta} + \frac{\sin\varphi}{\tan\vartheta}\partial_{\varphi}\,,\\
R_3 &= -\partial_{\varphi}\,,
\end{align}
\end{subequations}
as well as the translation generators
\begin{subequations}\label{eq:gentra}
\begin{align}
T_1 &= \chi\sin\vartheta\cos\varphi\partial_r + \frac{\chi}{r}\cos\vartheta\cos\varphi\partial_{\vartheta} - \frac{\chi\sin\varphi}{r\sin\vartheta}\partial_{\varphi}\,,\\
T_2 &= \chi\sin\vartheta\sin\varphi\partial_r + \frac{\chi}{r}\cos\vartheta\sin\varphi\partial_{\vartheta} + \frac{\chi\cos\varphi}{r\sin\vartheta}\partial_{\varphi}\,,\\
T_3 &= \chi\cos\vartheta\partial_r - \frac{\chi}{r}\sin\vartheta\partial_{\vartheta}\,.
\end{align}
\end{subequations}
Here we used the abbreviation \(\chi = \sqrt{1 - u^2r^2}\), where \(u\) is a constant parameter, which may take any real or imaginary value. A more conventional approach is to define the spatial curvature parameter \(k = u^2\), and to restrict its values to \(k \in \{-1,0,1\}\). However, for our purposes it will turn out to be more convenient to work with general real or imaginary \(u\) instead. It follows that the Lie brackets of the generators are given by
\begin{equation}
[R_i, R_j] = \epsilon_{ijk}R_k\,, \quad
[T_i, T_j] = u^2\epsilon_{ijk}R_k\,, \quad
[T_i, R_j] = \epsilon_{ijk}T_k\,.
\end{equation}
They generate the (connected) symmetry groups \(\mathrm{SO}(4)\) for \(u^2 > 0\), \(\mathrm{ISO}(3)\) for \(u^2 = 0\) and \(\mathrm{SO}_0(1,3)\) for \(u^2 < 0\), where the latter denotes the proper orthochronous Lorentz group, i.e., the connected component of the unit element in \(\mathrm{SO}(1,3)\). We will extend our consideration to non-connected groups by considering reflections in section~\ref{ssec:ctreflect}.

\subsection{Lorentz algebra}\label{ssec:loralgeb}
For later use we also introduce the notation and convention we employ for the Lorentz algebra \(\mathfrak{so}(1,3)\). A basis is given by the generators \(J_i\) of rotations and \(K_i\) of boosts, where \(i = 1, \ldots, 3\). They satisfy the commutation relations
\begin{equation}
[J_i, J_j] = \epsilon_{ijk}J_k\,, \quad
[K_i, K_j] = -\epsilon_{ijk}J_k\,, \quad
[K_i, J_j] = \epsilon_{ijk}K_k\,.
\end{equation}
An explicit matrix representation is given by
\begin{subequations}\label{eq:loralgmat}
\begin{align}
J_1 &= \begin{pmatrix}
0 & 0 & 0 & 0\\
0 & 0 & 0 & 0\\
0 & 0 & 0 & -1\\
0 & 0 & 1 & 0
\end{pmatrix}\,, &
J_2 &= \begin{pmatrix}
0 & 0 & 0 & 0\\
0 & 0 & 0 & 1\\
0 & 0 & 0 & 0\\
0 & -1 & 0 & 0
\end{pmatrix}\,, &
J_3 &= \begin{pmatrix}
0 & 0 & 0 & 0\\
0 & 0 & -1 & 0\\
0 & 1 & 0 & 0\\
0 & 0 & 0 & 0
\end{pmatrix}\,,\\
K_1 &= \begin{pmatrix}
0 & 1 & 0 & 0\\
1 & 0 & 0 & 0\\
0 & 0 & 0 & 0\\
0 & 0 & 0 & 0
\end{pmatrix}\,, &
K_2 &= \begin{pmatrix}
0 & 0 & 1 & 0\\
0 & 0 & 0 & 0\\
1 & 0 & 0 & 0\\
0 & 0 & 0 & 0
\end{pmatrix}\,, &
K_3 &= \begin{pmatrix}
0 & 0 & 0 & 1\\
0 & 0 & 0 & 0\\
0 & 0 & 0 & 0\\
1 & 0 & 0 & 0
\end{pmatrix}\,.
\end{align}
\end{subequations}
We will make use of these expressions when we calculate the symmetric tetrads in the following section.

\subsection{Complexification and pseudo-tetrads}\label{ssec:complex}
As discussed in section~\ref{ssec:stsym}, the notion of spacetime symmetry for a teleparallel geometry, formulated in terms of the tetrad in the Weitzenböck gauge, involves the choice of a homomorphism \(\boldsymbol{\Lambda}: G \to \mathrm{SO}(1,3)\) from the symmetry group to the Lorentz group. Representing the latter by its natural representation displayed in the preceding section, \(\boldsymbol{\Lambda}\) in particular constitutes a four-dimensional, real representation of the symmetry group. However, during the later sections of this article, it will turn out to be more convenient to consider complex representations instead, as this approach will allow a unified description of otherwise disconnected branches of solutions. This comes with the drawback that the tetrads and spin connections we construct will, in general, also be complex, and so should more appropriately be denoted ``pseudo-tetrads''. This does not pose any difficulties for the notion of symmetry we use, since all definitions and equations used in section~\ref{ssec:stsym} also apply for complex-valued tensor fields, but of course, any physically observable fields, as well as the action of any physical theory, must be real, and so we will regard as physical only real tetrads and spin connections.

We finally remark that complex tetrads and spin connections have been considered for the description of teleparallel cosmology in earlier works. An explicit form for a complex tetrad obeying cosmological symmetry has been given in hyperspherical coordinates in~\cite{Ferraro:2011us,Capozziello:2018hly} and in spherical coordinates in~\cite{Hohmann:2019nat}.

\section{Construction of cosmologically symmetric teleparallel geometry}\label{sec:construction}
We will now explicitly construct the most general class of teleparallel geometries obeying the cosmological symmetry, as detailed in the previous section. This can be done using different approaches, and for illustrative purposes we will demonstrate three different approaches, and show that they lead to the same result, thus proving their consistency. In section~\ref{ssec:algebra}, we will make use of the notion of symmetry for a teleparallel geometry represented by a tetrad and spin connection as detailed in section~\ref{ssec:stsym}, by constructing all possible homomorphisms from the group of spacetime symmetries into the Lorentz group. This approach is the most direct, as it yields explicit expressions for the tetrad and spin connection. The other two approaches take as their starting points the metric-affine representation of teleparallel geometry. In section~\ref{ssec:metaffine}, we start from the most general cosmologically symmetric metric-affine geometry, and successively apply the conditions of metricity and flatness in order to obtain teleparallel geometries. A related approach is presented in section~\ref{ssec:decomp}, which makes use of the fact that the affine connection of a teleparallel geometry is fully characterized by its torsion, and that the latter may be decomposed into three irreducible parts under the Lorentz group.

\subsection{Lorentz algebra and representation approach}\label{ssec:algebra}
The first approach we present here is the most explicit one, as it will directly yield the tetrads and spin connections which define the cosmologically symmetric teleparallel geometry. For this purpose, it makes use of the notion of symmetry detailed in section~\ref{ssec:stsym}. Its application requires to construct homomorphisms \(\boldsymbol{\Lambda}: G \to \mathrm{SO}(1,3)\) from the spacetime symmetry group to the Lorentz group (or, more generally, its complexification). Here we construct \emph{all} such homomorphisms in the case of cosmological symmetry, by proceeding in three steps. First, we list all irreducible, at most four-dimensional representations of the spacetime symmetry groups in section~\ref{sssec:arirrep}. From these we construct all four-dimensional (complex) representations of the symmetry groups in section~\ref{sssec:ar4rep}. As shown finally in section~\ref{sssec:arlorrep}, it will turn out that each of these representations yields a unique homomorphism into the complexified Lorentz group. We then use the obtained homomorphisms to construct all symmetric tetrads in the Weitzenböck gauge in section~\ref{sssec:arwbtet}. For convenience, we also transform these to the diagonal gauge in section~\ref{sssec:ardiagtet}.

\subsubsection{Irreducible representations of the symmetry group}\label{sssec:arirrep}
The first step of our construction is to find all irreducible representations of the spacetime symmetry groups listed in section~\ref{ssec:cosmosym} whose dimension is at most four, so that they can serve as building blocks in order to construct four-dimensional representations in the next step. For the groups we encounter in cosmology, the irreducible representations are completely classified, and so it will be sufficient to list the relevant representations. As argued in section~\ref{ssec:complex}, it is most convenient to consider representations on a complex vector space, and to consider those which admit a restriction to a real vector space later. Following our discussion in section~\ref{ssec:cosmosym}, we consider three cases, where the latter two can be grouped together:
\begin{enumerate}
\item
For \(u = 0\), the symmetry group is given by the Euclidean group \(\mathrm{ISO}(3)\). Its finite-dimensional irreducible representations are uniquely obtained by applying the method of induced representations to the corresponding irreducible representations of the rotation group \(\mathrm{SO}(3)\). The latter are obtained by considering the (complex) representations of its Lie algebra \(\mathfrak{so}(3) \cong \mathfrak{su}(2)\), which enumerated by their integer or half-integer spin \(2s \in \mathbb{N}\) and have dimension \(2s + 1\). Hence, restricting to at most four-dimensional representations, we find the restriction \(0 \leq s \leq 3/2\). However, recalling that only the integer spin representations of odd dimension can be integrated to group representations of \(\mathrm{SO}(3)\), it follows that the only possible choices are \(s = 0\) and \(s = 1\). Explicitly, we will use these representations in the following form, acting on the generators of the spacetime symmetry algebra:
\begin{enumerate}
\item
Trivial (scalar) representation \(\mathbf{0}\): both rotation and translation generators \(R_i\) and \(T_i\) are mapped to \(\mathds{0}_1 \in \mathfrak{gl}(1, \mathbb{C})\).
\item
Vector representation \(\mathbf{1}\): rotation generators are mapped via
\begin{equation}\label{eq:flatvecrep}
R_1 \mapsto \begin{pmatrix}
0 & 0 & 0\\
0 & 0 & -1\\
0 & 1 & 0
\end{pmatrix}\,, \quad
R_2 \mapsto \begin{pmatrix}
0 & 0 & 1\\
0 & 0 & 0\\
-1 & 0 & 0
\end{pmatrix}\,, \quad
R_3 \mapsto \begin{pmatrix}
0 & -1 & 0\\
1 & 0 & 0\\
0 & 0 & 0
\end{pmatrix}\,,
\end{equation}
while translation generators \(T_i\) are mapped to the zero element \(\mathds{0}_3 \in \mathfrak{gl}(3, \mathbb{C})\).
\end{enumerate}

\item
In the cases \(u^2 > 0\), with symmetry group \(\mathrm{SO}(4)\) and Lie algebra \(\mathfrak{so}(4) \cong \mathfrak{su}(2) \oplus \mathfrak{su}(2)\), and \(u^2 < 0\), with symmetry group \(\mathrm{SO}^+(1,3)\) and Lie algebra \(\mathfrak{so}(1,3) \cong \mathfrak{sl}(2, \mathbb{C})\), the irreducible representations are most easily obtained from their common complexification \(\mathfrak{sl}(2, \mathbb{C}) \oplus \mathfrak{sl}(2, \mathbb{C})\). Its (complex linear) representations are labeled by pairs \((m,n)\) of non-negative integers or half-integers, \(2m \in \mathbb{N}\) and \(2n \in \mathbb{N}\), with dimension \((2m + 1)(2n + 1)\). They can be integrated to representations of the relevant groups if and only if \(m + n \in \mathbb{N}\). Hence, up to dimension four, we find four irreducible representations \((0,0), \left(\frac{1}{2},\frac{1}{2}\right), (1,0), (0,1)\). Explicit forms of the representation matrices can be obtained, e.g., by using the spin matrices known from quantum mechanics. In the remaining sections, we will use the following conventions:
\begin{enumerate}
\item
Trivial representation \((0,0)\): as in the case \(u = 0\), both rotation and translation generators \(R_i\) and \(T_i\) are mapped to \(\mathds{0}_1 \in \mathfrak{gl}(1, \mathbb{C})\).

\item
Vector representation \(\left(\frac{1}{2},\frac{1}{2}\right)\):
\begin{subequations}\label{eq:curvvecrep}
\begin{align}
\hspace*{-7.5mm}R_1 &\mapsto -\frac{i}{2}\begin{pmatrix}
0 & 1 & 1 & 0\\
1 & 0 & 0 & 1\\
1 & 0 & 0 & 1\\
0 & 1 & 1 & 0
\end{pmatrix}\,, &
R_2 &\mapsto \frac{1}{2}\begin{pmatrix}
0 & 1 & 1 & 0\\
-1 & 0 & 0 & 1\\
-1 & 0 & 0 & 1\\
0 & -1 & -1 & 0
\end{pmatrix}\,, &
R_3 &\mapsto \begin{pmatrix}
i & 0 & 0 & 0\\
0 & 0 & 0 & 0\\
0 & 0 & 0 & 0\\
0 & 0 & 0 & -i
\end{pmatrix}\,,\\
\hspace*{-7.5mm}T_1 &\mapsto \frac{iu}{2}\begin{pmatrix}
0 & -1 & 1 & 0\\
-1 & 0 & 0 & 1\\
1 & 0 & 0 & -1\\
0 & 1 & -1 & 0
\end{pmatrix}\,, &
T_2 &\mapsto \frac{u}{2}\begin{pmatrix}
0 & 1 & -1 & 0\\
-1 & 0 & 0 & -1\\
1 & 0 & 0 & 1\\
0 & 1 & -1 & 0
\end{pmatrix}\,, &
T_3 &\mapsto \begin{pmatrix}
0 & 0 & 0 & 0\\
0 & -iu & 0 & 0\\
0 & 0 & iu & 0\\
0 & 0 & 0 & 0
\end{pmatrix}\,.
\end{align}
\end{subequations}

\item
Self-dual and anti-self-dual two-form representations \((1,0)\) (upper sign) and \((0,1)\) (lower sign):
\begin{subequations}\label{eq:curvsdfrep}
\begin{align}
R_1 &\mapsto -\frac{i}{\sqrt{2}}\begin{pmatrix}
0 & 1 & 0\\
1 & 0 & 1\\
0 & 1 & 0
\end{pmatrix}\,, &
R_2 &\mapsto \frac{1}{\sqrt{2}}\begin{pmatrix}
0 & 1 & 0\\
-1 & 0 & 1\\
0 & -1 & 0
\end{pmatrix}\,, &
R_3 &\mapsto \begin{pmatrix}
i & 0 & 0\\
0 & 0 & 0\\
0 & 0 & -i
\end{pmatrix}\,,\\
T_1 &\mapsto \mp\frac{iu}{\sqrt{2}}\begin{pmatrix}
0 & 1 & 0\\
1 & 0 & 1\\
0 & 1 & 0
\end{pmatrix}\,, &
T_2 &\mapsto \pm\frac{u}{\sqrt{2}}\begin{pmatrix}
0 & 1 & 0\\
-1 & 0 & 1\\
0 & -1 & 0
\end{pmatrix}\,, &
T_3 &\mapsto \pm\begin{pmatrix}
iu & 0 & 0\\
0 & 0 & 0\\
0 & 0 & -iu
\end{pmatrix}\,.
\end{align}
\end{subequations}
\end{enumerate}
\end{enumerate}
We will make use of these explicit matrix expressions in the following sections, as they enter the derivation of the cosmologically symmetric tetrads.

\subsubsection{Four-dimensional representations of the symmetry group}\label{sssec:ar4rep}
We can now use the irreducible representations listed in the previous section in order to construct all four-dimensional representations of the relevant symmetry groups. They are given by the following direct sums:
\begin{enumerate}
\item
In the case \(u = 0\) we have two possibilities:
\begin{enumerate}
\item
The trivial representation: \(\boldsymbol{\Phi} = \mathbf{0} \oplus \mathbf{0} \oplus \mathbf{0} \oplus \mathbf{0}\).
\item
The vector representation: \(\boldsymbol{\Phi} = \mathbf{0} \oplus \mathbf{1}\).
\end{enumerate}
\item
For \(u \neq 0\), there are four representations:
\begin{enumerate}
\item
The trivial representation: \(\boldsymbol{\Phi} = (0,0) \oplus (0,0) \oplus (0,0) \oplus (0,0)\).
\item
The vector representation: \(\boldsymbol{\Phi} = \left(\frac{1}{2}, \frac{1}{2}\right)\).
\item
The self-dual two-form representation: \(\boldsymbol{\Phi} = (0,0) \oplus (1,0)\).
\item
The anti-self-dual two-form representation: \(\boldsymbol{\Phi} = (0,0) \oplus (0,1)\).
\end{enumerate}
\end{enumerate}
The construction of the block-diagonal representation matrices is straightforward, and so we will omit their explicit forms here for brevity.

\subsubsection{Preservation of the Minkowski metric}\label{sssec:arlorrep}
Having determined all (complex) four-dimensional representations of the cosmological symmetry groups, we must isolate those representations whose image lies inside the (complexified) Lorentz group
\begin{equation}
\mathrm{SO}(1, 3, \mathbb{C}) = \{\Lambda \in \mathrm{GL}(4, \mathbb{C}), \Lambda^t\eta\Lambda = \eta\} \cong \mathrm{SO}(4, \mathbb{C})\,,
\end{equation}
where the latter relation stems from the fact that in the complex orthogonal case, the signature of the metric becomes irrelevant due to the presence of the imaginary unit. Denoting by \(\boldsymbol{\Phi}\) any of the four-dimensional representations found in the preceding section, we thus need to find a transformation matrix \(P \in \mathrm{GL}(4, \mathbb{C})\) such that
\begin{equation}~\label{eq:lorbastrans}
\boldsymbol{\Lambda}_u = P^{-1}\boldsymbol{\Phi}_uP \in \mathrm{SO}(4, \mathbb{C})
\end{equation}
for all \(u \in G\). Hence, by definition, \(P\) must satisfy
\begin{equation}
\eta = \boldsymbol{\Lambda}_u^t\eta\boldsymbol{\Lambda}_u = P^t\boldsymbol{\Phi}_u^tP^{-1\,t}\eta P^{-1}\boldsymbol{\Phi}_uP\,.
\end{equation}
It will turn out to be easier to work with the Lie algebra representations instead. Denoting by \(\boldsymbol{\phi}\) and \(\boldsymbol{\lambda}\) the Lie algebra representations induced by \(\boldsymbol{\Phi}\) and \(\boldsymbol{\Lambda}\), respectively. In terms of these the condition on \(P\) reads
\begin{equation}
0 = \boldsymbol{\lambda}_{\xi}^t\eta + \eta\boldsymbol{\lambda}_{\xi} = P^t\boldsymbol{\phi}_{\xi}^tP^{-1\,t}\eta + \eta P^{-1}\boldsymbol{\phi}_{\xi}P\,.
\end{equation}
for all \(\xi \in \mathfrak{g}\). By a suitable multiplication of this equation by \(P\) and its transpose from either side we can eliminate the inverse of \(P\) and obtain the simpler equation
\begin{equation}
0 = P\eta P^t\boldsymbol{\phi}_{\xi}^t + \boldsymbol{\phi}_{\xi}P\eta P^t\,,
\end{equation}
which is now only quadratic in \(P\), but does not involve its inverse. Finally, note that it depends on the matrix \(\tilde{P} = P\eta P^t\), which is symmetric and invertible by construction, and that it is linear in this matrix. Writing the equation for \(\tilde{P}\) in the form
\begin{equation}
0 = \tilde{P}\boldsymbol{\phi}_{\xi}^t + \boldsymbol{\phi}_{\xi}\tilde{P}\,,
\end{equation}
we see that it simply represents the symmetric, non-degenerate bilinear form preserved by \(\boldsymbol{\Phi}\). Hence, we need to find the symmetric, invertible matrices \(\tilde{P}\) which satisfy this equation. This is a simple task, and it turns out to yield solutions for all representations we consider. In particular, we obtain the following bilinear forms.
\begin{enumerate}
\item
Trivial representation \(\boldsymbol{\Phi} = \mathbf{0} \oplus \mathbf{0} \oplus \mathbf{0} \oplus \mathbf{0}\) or \(\boldsymbol{\Phi} = (0,0) \oplus (0,0) \oplus (0,0) \oplus (0,0)\): In this case any non-degenerate bilinear form is trivially preserved. We may therefore choose \(P = \mathds{1}_4\), and hence \(\boldsymbol{\Lambda} = \boldsymbol{\Phi}\). Obviously, the induced mapping of the Lie algebra generators is the trivial one
\begin{equation}\label{eq:homtriv}
R_i \mapsto 0\,, \quad
T_i \mapsto 0\,.
\end{equation}

\item
In the case \(u = 0\), we also have the non-trivial vector representation \(\boldsymbol{\Phi} = \mathbf{0} \oplus \mathbf{1}\). Using the representation matrices~\eqref{eq:flatvecrep}, we find that they preserve any symmetric, bilinear form of the type
\begin{equation}
\tilde{P} = \mathrm{diag}(a, b, b, b)\,.
\end{equation}
This obviously includes the Minkowski metric \(a = -1, b = 1\), and so we may also in this case choose \(P = \mathds{1}_4\), and hence \(\boldsymbol{\Lambda} = \boldsymbol{\Phi}\). In terms of the generators~\eqref{eq:loralgmat} of the Lorentz algebra, we may thus write the homomorphism \(\boldsymbol{\Lambda}\) as
\begin{equation}\label{eq:homflat}
R_i \mapsto J_i\,, \quad
T_i \mapsto 0\,.
\end{equation}
Note that this is a real homomorphism.

\item
For \(u \neq 0\) and the vector representation \(\boldsymbol{\Phi} = \left(\frac{1}{2}, \frac{1}{2}\right)\), given in matrix form by the assignment~\eqref{eq:curvvecrep}, we find that the most general preserved symmetric bilinear form is given by
\begin{equation}
\tilde{P} = \begin{pmatrix}
0 & 0 & 0 & a\\
0 & 0 & -a & 0\\
0 & -a & 0 & 0\\
a & 0 & 0 & 0
\end{pmatrix}\,.
\end{equation}
There are different possible choices for the matrix \(P\) which satisfy \(\tilde{P} = P\eta P^t\) for some value of \(a\). Keeping in mind that the magnitude of \(a\) is not relevant, as it cancels in the basis transformation, we find that two such choices are given by the explicit form
\begin{equation}
P = \begin{pmatrix}
0 & 1 & i & 0\\
\mp 1 & 0 & 0 & 1\\
\pm 1 & 0 & 0 & 1\\
0 & -1 & i & 0
\end{pmatrix}\,,
\end{equation}
leading to the group homomorphism \(\boldsymbol{\Lambda}\), whose Lie algebra homomorphism \(\boldsymbol{\lambda}\) is given by
\begin{equation}\label{eq:homvec}
R_i \mapsto J_i\,, \quad
T_i \mapsto \pm iuK_i\,.
\end{equation}
Note that the two choices for the sign we show here are related by the time reflection, which is a (non-proper) Lorentz transformation sending \(K_i\) to \(-K_i\), but leaving \(J_i\) unchanged. Hence, they constitute equivalent representations. We list both choices here for later convenience. Also observe that these homomorphisms are real if and only if \(u^2 < 0\).

\item
Finally, the case \(u \neq 0\) admits the two remaining representations \(\boldsymbol{\Phi} = (0,0) \oplus (1,0)\) or \(\boldsymbol{\Phi} = (0,0) \oplus (0,1)\). It turns out that both of them, using the explicit representation matrices~\eqref{eq:curvsdfrep}, preserve the most general symmetric bilinear form
\begin{equation}
\tilde{P} = \begin{pmatrix}
a & 0 & 0 & 0\\
0 & 0 & 0 & b\\
0 & 0 & -b & 0\\
0 & b & 0 & 0
\end{pmatrix}\,.
\end{equation}
A possible basis transformation, which brings the Minkowski metric into this form and thus holds for both representations, is given by
\begin{equation}
P = \begin{pmatrix}
1 & 0 & 0 & 0\\
0 & -i & 1 & 0\\
0 & 0 & 0 & -i\sqrt{2}\\
0 & i & 1 & 0
\end{pmatrix}
\end{equation}
Hence, we find that the obtained homomorphisms can be expressed as
\begin{equation}\label{eq:homsdf}
R_i \mapsto J_i\,, \quad
T_i \mapsto \pm uJ_i\,,
\end{equation}
where the upper sign stands for \(\boldsymbol{\Phi} = (0,0) \oplus (0,1)\), while the lower sign stands for \(\boldsymbol{\Phi} = (0,0) \oplus (1,0)\). Note that in this case the homomorphisms are not equivalent, since they are not related by a Lorentz transformation. They are real if and only if \(u^2 > 0\).
\end{enumerate}
Note that instead of the Lie group homomorphisms and Lie algebra homomorphisms listed here, one could equivalently use any other choice of homomorphism which differs only by conjugation with another Lorentz transformation.

\subsubsection{Symmetric tetrads in the Weitzenböck gauge}\label{sssec:arwbtet}
We can now use the homomorphisms found in the previous section and derive solutions to the symmetry condition~\eqref{eq:tetinfsym} in the Weitzenböck gauge. Due to the structure of the homomorphisms we found, it is most useful to proceed in two steps. First, we determine the most general tetrad in the Weitzenböck gauge which obeys spherical symmetry. For this step it is sufficient to consider only the rotation generators. Note that following our discussion in the previous section, the homomorphisms split into two classes, either mapping the rotation generators \(R_i\) to their counterparts \(J_i\) in the Lorentz algebra, or to the zero elements. For both choices, the most general symmetric tetrads are known~\cite{Hohmann:2019nat}:
\begin{enumerate}
\item
In the trivial case \(R_i \mapsto 0\), no symmetric tetrads exist due to a topological obstruction. This follows from the fact that the orbits of the rotation group are homeomorphic to the sphere \(S^2\), and that a symmetric tetrad would in this case allow to construct a parallelization of \(S^2\). However, \(S^2\) is not parallelizable, leading to a contradiction. See~\cite{Hohmann:2019nat} for a detailed derivation.

\item
In the non-trivial case \(R_i \mapsto J_i\) the most general symmetric tetrad reads~\cite{Hohmann:2019nat,Hohmann:2019fvf}
\begin{subequations}\label{eq:sphertetradwb}
\begin{align}
\theta^0 &= \mathcal{F}_1\cosh\mathcal{F}_3\dd t + \mathcal{F}_2\sinh\mathcal{F}_4\dd r\,,\\
\theta^1 &= \sin\vartheta\cos\varphi(\mathcal{F}_1\sinh\mathcal{F}_3\dd t + \mathcal{F}_2\cosh\mathcal{F}_4\dd r)\nonumber\\
&\phantom{=}+ \mathcal{F}_5\big[(\cos\mathcal{F}_6\cos\vartheta\cos\varphi - \sin\mathcal{F}_6\sin\varphi)\dd\vartheta\nonumber\\
&\phantom{=}\quad - \sin\vartheta(\cos\mathcal{F}_6\sin\varphi + \sin\mathcal{F}_6\cos\vartheta\cos\varphi)\dd\varphi\big]\,,\\
\theta^2 &= \sin\vartheta\sin\varphi(\mathcal{F}_1\sinh\mathcal{F}_3\dd t + \mathcal{F}_2\cosh\mathcal{F}_4\dd r)\nonumber\\
&\phantom{=}+ \mathcal{F}_5\big[(\cos\mathcal{F}_6\cos\vartheta\sin\varphi + \sin\mathcal{F}_6\cos\varphi)\dd\vartheta\nonumber\\
&\phantom{=}\quad + \sin\vartheta(\cos\mathcal{F}_6\cos\varphi - \sin\mathcal{F}_6\cos\vartheta\sin\varphi)\dd\varphi\big]\,,\\
\theta^3 &= \cos\vartheta(\mathcal{F}_1\sinh\mathcal{F}_3\dd t + \mathcal{F}_2\cosh\mathcal{F}_4\dd r)\nonumber\\
&\phantom{=}+ \mathcal{F}_5\left[-\cos\mathcal{F}_6\sin\vartheta\dd\vartheta + \sin\mathcal{F}_6\sin^2\vartheta\dd\varphi\right]\,,
\end{align}
\end{subequations}
and is thus parametrized by six functions \(\mathcal{F}_1(t, r), \ldots, \mathcal{F}_6(t, r)\) of the two coordinates \(t, r\).
\end{enumerate}
Hence, we see that for the trivial homomorphism~\eqref{eq:homtriv} no symmetric tetrads exist, and we are left with the non-trivial homomorphisms. We thus use the translation generators in order to impose cosmological symmetry. Depending on the choice of the homomorphism, we find the following results:
\begin{enumerate}
\item
For the vector case \(T_i \mapsto \pm iuK_i\), we find the solution
\begin{equation}
\hspace*{-3mm}\mathcal{F}_1 = \mathcal{N}\,, \quad
\mathcal{F}_2 = \frac{\mathcal{A}}{\chi}\,, \quad
\mathcal{F}_3 = \mathcal{F}_4 = \pm\arcsinh(iur)\,, \quad
\mathcal{F}_5 = \mathcal{A}r\,, \quad
\mathcal{F}_6 = 0\,,
\end{equation}
so that the tetrad reads
\begin{subequations}\label{eq:tetwbvec}
\begin{align}
\theta^0 &= \mathcal{N}\chi\dd t \pm iu\mathcal{A}\frac{r}{\chi}\dd r\,,\\
\theta^1 &= \mathcal{A}\left[\sin\vartheta\cos\varphi\left(\dd r \pm iu\frac{\mathcal{N}}{\mathcal{A}}r\dd t\right) + r\cos\vartheta\cos\varphi\dd\vartheta - r\sin\vartheta\sin\varphi\dd\varphi\right]\,,\\
\theta^2 &= \mathcal{A}\left[\sin\vartheta\sin\varphi\left(\dd r \pm iu\frac{\mathcal{N}}{\mathcal{A}}r\dd t\right) + r\cos\vartheta\sin\varphi\dd\vartheta + r\sin\vartheta\cos\varphi\dd\varphi\right]\,,\\
\theta^3 &= \mathcal{A}\left[\cos\vartheta\left(\dd r \pm iu\frac{\mathcal{N}}{\mathcal{A}}r\dd t\right) - r\sin\vartheta\dd\vartheta\right]\,.
\end{align}
\end{subequations}
Note that the two different sign choices can be absorbed by changing the sign of either \(\mathcal{N}\) or \(t\), as well as performing a Lorentz transformation \(\Lambda = \mathrm{diag}(-1,1,1,1)\) changing the sign of the tetrad component \(\theta^0\), and so these are equivalent geometries.
\item
For the two-form case \(T_i \mapsto \pm uJ_i\), a solution is given by
\begin{equation}
\mathcal{F}_1 = \mathcal{N}\,, \quad
\mathcal{F}_2 = \frac{\mathcal{A}}{\chi}\,, \quad
\mathcal{F}_5 = \mathcal{A}r\,, \quad
\mathcal{F}_3 = \mathcal{F}_4 = 0\,, \quad
\mathcal{F}_6 = \mp\arcsin(ur)\,,
\end{equation}
and we find the tetrad
\begin{subequations}\label{eq:tetwbsdf}
\begin{align}
\theta^0 &= \mathcal{N}\dd t\,,\\
\theta^1 &= \mathcal{A}\bigg[\frac{\sin\vartheta\cos\varphi}{\chi}\dd r + r(\chi\cos\vartheta\cos\varphi \pm ur\sin\varphi)\dd\vartheta\nonumber\\
&\phantom{= \mathcal{A}\bigg[} \quad - r\sin\vartheta(\chi\sin\varphi \mp ur\cos\vartheta\cos\varphi)\dd\varphi\bigg]\,,\\
\theta^2 &= \mathcal{A}\bigg[\frac{\sin\vartheta\sin\varphi}{\chi}\dd r + r(\chi\cos\vartheta\sin\varphi \mp ur\cos\varphi)\dd\vartheta\nonumber\\
&\phantom{= \mathcal{A}\bigg[} \quad + r\sin\vartheta(\chi\cos\varphi \pm ur\cos\vartheta\sin\varphi)\dd\varphi\bigg]\,,\\
\theta^3 &= \mathcal{A}\left[\frac{\cos\vartheta}{\chi}\dd r - r\chi\sin\vartheta\dd\vartheta \mp ur^2\sin^2\vartheta\dd\varphi\right]\,.
\end{align}
\end{subequations}
In this case the different signs correspond to inequivalent geometries, since the two tetrads are not related by a Lorentz transformation and change of parametrization.
\end{enumerate}
Here \(\mathcal{N} = \mathcal{N}(t)\) and \(\mathcal{A} = \mathcal{A}(t)\) are free functions of the cosmological time \(t\) which parametrize the solutions.

\subsubsection{Symmetric tetrads in the diagonal gauge}\label{sssec:ardiagtet}
Finally, we transform the tetrads we have determined in the Weitzenböck gauge in the previous section into the diagonal gauge. Observe that both tetrads~\eqref{eq:tetwbvec} and~\eqref{eq:tetwbsdf} have a common diagonal form
\begin{equation}\label{eq:tetdiag}
\theta'^0 = \mathcal{N}\dd t\,, \quad
\theta'^1 = \frac{\mathcal{A}}{\chi}\dd r\,, \quad
\theta'^2 = \mathcal{A}r\dd\vartheta\,, \quad
\theta'^3 = \mathcal{A}r\sin\vartheta\dd\varphi\,,
\end{equation}
where the parameter \(u\) defining the spacetime symmetry group enters only through the cosmological distance function \(\chi\). Here the prime \(\prime\) indicates the diagonal gauge. It is now straightforward to calculate the Lorentz transformations \(\Lambda^A{}_B\) which relate the Weitzenböck and diagonal gauges by solving the relation
\begin{equation}
\theta'^A{}_{\mu} = \Lambda^A{}_B\theta^B{}_{\mu}\,,
\end{equation}
and to calculate the corresponding spin connections
\begin{equation}
\omega'^A{}_{B\mu} = \Lambda^A{}_C\partial_{\mu}(\Lambda^{-1})^C{}_B\,.
\end{equation}
We find the following Lorentz transformations and spin connections:
\begin{enumerate}
\item
For the tetrads~\eqref{eq:tetwbvec} obtained from the vector representation, we find the Lorentz transformations
\begin{equation}\label{eq:lortrvec}
\Lambda^A{}_B = \begin{pmatrix}
\chi & \mp iur\sin\vartheta\cos\varphi & \mp iur\sin\vartheta\sin\varphi & \mp iur\cos\vartheta\\
\mp iur & \chi\sin\vartheta\cos\varphi & \chi\sin\vartheta\sin\varphi & \chi\cos\vartheta\\
0 & \cos\vartheta\cos\varphi & \cos\vartheta\sin\varphi & -\sin\vartheta\\
0 & -\sin\varphi & \cos\varphi & 0
\end{pmatrix}\,.
\end{equation}
The non-vanishing coefficients of the corresponding spin connections read
\begin{gather}
\hspace*{-5mm}
\omega'^1{}_{2\vartheta} = -\omega'^2{}_{1\vartheta} = -\chi\,, \quad
\omega'^1{}_{3\varphi}   = -\omega'^3{}_{1\varphi} = -\chi\sin\vartheta\,, \quad
\omega'^2{}_{3\varphi}   = -\omega'^3{}_{2\varphi} = -\cos\vartheta\,,\nonumber\\
\omega'^0{}_{1r}         = \omega'^1{}_{0r} = \mp\frac{iu}{\chi}\,, \quad
\omega'^0{}_{2\vartheta} = \omega'^2{}_{0\vartheta} = \mp iur\,, \quad
\omega'^0{}_{3\varphi}   = \omega'^3{}_{0\varphi} = \mp iur\sin\vartheta\,.\label{eq:spconvec}
\end{gather}
Note that changing the sign in the solution has two effects here. Recall that for the tetrad~\eqref{eq:tetwbvec} it corresponds to a combination of a global Lorentz transformation, which changes the sign of \(\theta^0\), together with a change of sign either of the lapse function \(\mathcal{N}\) or the time coordinate \(t\). While the former, being a global Lorentz transformation, does not affect the spin connection, the latter changes the sign of the basis vector \(\theta'^0\) in the diagonal tetrad~\eqref{eq:tetdiag}, and hence the sign of all components involving a $0$-index in the spin connection~\eqref{eq:spconvec}.

\item
For the tetrads~\eqref{eq:tetwbsdf} obtained from the two-form representations, the diagonalizing Lorentz transformations are given by
\begin{equation}\label{eq:lortrsdf}
\Lambda^A{}_B = \begin{pmatrix}
1 & 0 & 0 & 0\\
0 & \sin\vartheta\cos\varphi & \sin\vartheta\sin\varphi & \cos\vartheta\\
0 & \chi\cos\vartheta\cos\varphi \pm ur\sin\varphi & \chi\cos\vartheta\sin\varphi \mp ur\cos\varphi & -\chi\sin\vartheta\\
0 & \pm ur\cos\vartheta\cos\varphi - \chi\sin\varphi & \chi\cos\varphi \pm ur\cos\vartheta\sin\varphi & \mp ur\sin\vartheta
\end{pmatrix}\,.
\end{equation}
This yields the non-vanishing spin connection coefficients
\begin{gather}
\hspace*{-5mm}
\omega'^1{}_{3\varphi}   = -\omega'^3{}_{1\varphi} = -\chi\sin\vartheta\,, \quad
\omega'^2{}_{3r}         = -\omega'^3{}_{2r} = \pm\frac{u}{\chi}\,, \quad
\omega'^2{}_{3\varphi}   = -\omega'^3{}_{2\varphi} = -\cos\vartheta\,,\nonumber\\
\hspace*{-2.5mm}
\omega'^1{}_{2\vartheta} = -\omega'^2{}_{1\vartheta} = -\chi\,, \quad
\omega'^1{}_{2\varphi}   = -\omega'^2{}_{1\varphi} = \pm ur\sin\vartheta\,, \quad
\omega'^1{}_{3\vartheta} = -\omega'^3{}_{1\vartheta} = \mp ur\,.\label{eq:spconsdf}
\end{gather}
\end{enumerate}
One could now use the obtained tetrads and calculate further geometric quantities such as the coefficients \(\Gamma^{\mu}{}_{\nu\rho}\) of the affine connection, its torsion and contortion. However, it is instructive to proceed differently here. Instead of calculating the affine connection from the tetrads, we will show in the following sections that it can also be obtained directly by imposing cosmological symmetry on a metric-affine geometry, and establish teleparallelism after this step, and that this approach fully recovers the tetrads we have found above. We will use these independent approaches to obtain the connection coefficients in section~\ref{ssec:metaffine} and the torsion and contortion in section~\ref{ssec:decomp} below.

\subsection{Metric-affine approach}\label{ssec:metaffine}
The second approach we consider may be coined the ``metric-affine approach'', since it consists in first determining the most general metric-affine geometry with cosmological symmetry, and then further restricting this geometry by imposing the teleparallel conditions. We outline this construction in the following steps. In section~\ref{sssec:macosmosym}, we review the most general cosmologically symmetric metric-affine geometry, and introduce a few helpful quantities which will be used in the remaining sections. We then impose the conditions of metricity in section~\ref{sssec:mametriccomp} and flatness in section~\ref{sssec:maflat}. This will yield us the teleparallel affine connection. To obtain the tetrad and spin connection, we first choose a diagonal tetrad in section~\ref{sssec:madiagtet}, whose spin connection we determine by imposing the tetrad postulate, and then transform the result into the Weitzenböck gauge in section~\ref{sssec:mawbtet}.

\subsubsection{Cosmologically symmetric metric-affine geometry}\label{sssec:macosmosym}
It is well known that the most general metric of Lorentzian signature which satisfies the cosmological symmetry outlined in section~\ref{ssec:cosmosym} is the Robertson-Walker metric
\begin{equation}\label{eq:metcosmo}
g_{tt} = -\mathcal{N}^2\,, \quad
g_{rr} = \frac{\mathcal{A}^2}{\chi^2}\,, \quad
g_{\vartheta\vartheta} = \mathcal{A}^2r^2\,, \quad
g_{\varphi\varphi} = g_{\vartheta\vartheta}\sin^2\vartheta\,,
\end{equation}
which is parametrized by the lapse \(\mathcal{N}(t)\) and scale factor \(\mathcal{A}(t)\). It allows us to introduce a few helpful quantities, which will be useful in the remaining derivation. First, note that together with the choice of an orientation, the metric defines a volume form expressed by the totally antisymmetric Levi-Civita tensor with components
\begin{equation}
\epsilon_{tr\vartheta\varphi} = \frac{\mathcal{N}\mathcal{A}^3r^2\sin\vartheta}{\chi}\,, \quad
\epsilon^{tr\vartheta\varphi} = -\frac{\chi}{\mathcal{N}\mathcal{A}^3r^2\sin\vartheta}\,.
\end{equation}
Further, we denote by \(N\) the future pointing, unit vector field normal to the spatial hypersurfaces, and by \(n\) its dual covector field, given by
\begin{equation}
N = N^{\mu}\partial_{\mu} = \frac{1}{\mathcal{N}}\partial_t\,, \quad
n = n_{\mu}\dd x^{\mu} = -\mathcal{N}\dd t\,.
\end{equation}
They satisfy the relations
\begin{equation}
N \intprod n = g(N, N) = g^{-1}(n, n) = -1\,,
\end{equation}
and their components are identical up to raising and lowering indices with the metric, \(N^{\mu} = n^{\mu}\) and \(N_{\mu} = n_{\mu}\). On the spatial hypersurfaces we have the induced metric
\begin{equation}
h = g + n \otimes n = \mathcal{A}^2\left[\frac{\dd r \otimes \dd r}{\chi^2} + r^2(\dd\vartheta \otimes \dd\vartheta + \sin^2\vartheta\dd\varphi \otimes \dd\varphi)\right]\,.
\end{equation}
Together with the chosen orientation and time orientation, we can construct the volume form on the hypersurfaces in terms of the spatial Levi-Civita tensor
\begin{equation}
\varepsilon_{\mu\nu\rho} = n^{\sigma}\epsilon_{\sigma\mu\nu\rho}\,, \quad
\epsilon_{\mu\nu\rho\sigma} = 4\varepsilon_{[\mu\nu\rho}n_{\sigma]}\,.
\end{equation}
Its components are given by
\begin{equation}
\varepsilon_{r\vartheta\varphi} = \frac{\mathcal{A}^3r^2\sin\vartheta}{\chi}\,, \quad
\varepsilon^{r\vartheta\varphi} = \frac{\chi}{\mathcal{A}^3r^2\sin\vartheta}\,,
\end{equation}
while the remaining components are determined by its antisymmetry.

In addition to the metric we have the most general affine connection with cosmological symmetry, which can be written in the form~\cite{Hohmann:2019fvf,Minkevich:1998cv}
\begin{gather}
\Gamma^t{}_{tt} = \mathcal{K}_1\,, \quad
\Gamma^t{}_{rr} = \frac{\mathcal{K}_2}{\chi^2}\,, \quad
\Gamma^t{}_{\vartheta\vartheta} = \mathcal{K}_2r^2\,, \quad
\Gamma^t{}_{\varphi\varphi} = \mathcal{K}_2r^2\sin^2\vartheta\,,\nonumber\\
\Gamma^{\vartheta}{}_{r\vartheta} = \Gamma^{\vartheta}{}_{\vartheta r} = \Gamma^{\varphi}{}_{r\varphi} = \Gamma^{\varphi}{}_{\varphi r} = \frac{1}{r}\,, \quad
\Gamma^{\varphi}{}_{\vartheta\varphi} = \Gamma^{\varphi}{}_{\varphi\vartheta} = \cot\vartheta\,, \quad
\Gamma^{\vartheta}{}_{\varphi\varphi} = -\sin\vartheta\cos\vartheta\,,\nonumber\\
\Gamma^r{}_{\vartheta\vartheta} = -r\chi^2\,, \quad
\Gamma^r{}_{\varphi\varphi} = -r\chi^2\sin^2\vartheta\,, \quad
\Gamma^r{}_{\varphi\vartheta} = -\Gamma^r{}_{\vartheta\varphi} = \mathcal{K}_5r^2\chi\sin\vartheta\,,\nonumber\\
\Gamma^r{}_{tr} = \Gamma^{\vartheta}{}_{t\vartheta} = \Gamma^{\varphi}{}_{t\varphi} = \mathcal{K}_3\,, \quad
\Gamma^r{}_{rt} = \Gamma^{\vartheta}{}_{\vartheta t} = \Gamma^{\varphi}{}_{\varphi t} = \mathcal{K}_4\,, \quad
\Gamma^r{}_{rr} = \frac{u^2r}{\chi^2}\,,\nonumber\\
\Gamma^{\vartheta}{}_{r\varphi} = -\Gamma^{\vartheta}{}_{\varphi r} = \frac{\mathcal{K}_5\sin\vartheta}{\chi}\,, \quad
\Gamma^{\varphi}{}_{r\vartheta} = -\Gamma^{\varphi}{}_{\vartheta r} = -\frac{\mathcal{K}_5}{\chi\sin\vartheta}\,, \label{eq:affcosmo}
\end{gather}
and hence depends on five parameter functions \(\mathcal{K}_1(t), \ldots, \mathcal{K}_5(t)\).

\subsubsection{Metricity condition}\label{sssec:mametriccomp}
The next step in order to obtain a teleparallel geometry with cosmological symmetry is to impose the metricity condition~\eqref{eq:affnonmet} on the general cosmologically symmetric metric-affine geometry derived above. In terms of the unit normal covector field and spatial metric, we find that the nonmetricity is given by
\begin{equation}
Q_{\rho\mu\nu} = 2\mathcal{Q}_1n_{\rho}n_{\mu}n_{\nu} + 2\mathcal{Q}_2n_{\rho}h_{\mu\nu} + 2\mathcal{Q}_3h_{\rho(\mu}n_{\nu)}\,,
\end{equation}
where the three scalar quantities \(\mathcal{Q}_1(t), \ldots, \mathcal{Q}_3(t)\) are expressed in terms of the parameter functions in the metric and affine connection as
\begin{equation}
\mathcal{Q}_1 = \frac{\dot{\mathcal{N}}}{\mathcal{N}^2} - \frac{\mathcal{K}_1}{\mathcal{N}}\,, \quad
\mathcal{Q}_2 = \frac{1}{\mathcal{N}}\left(\mathcal{K}_4 - \frac{\dot{\mathcal{A}}}{\mathcal{A}}\right)\,, \quad
\mathcal{Q}_3 = \frac{\mathcal{K}_3}{\mathcal{N}} - \frac{\mathcal{K}_2\mathcal{N}}{\mathcal{A}^2}\,.
\end{equation}
It thus follows that the connection is metric-compatible, \(Q_{\mu\nu\rho} = 0\), if and only if
\begin{equation}\label{eq:metcompcosmo}
\mathcal{K}_1\mathcal{N} - \dot{\mathcal{N}} = \mathcal{K}_4\mathcal{A} - \dot{\mathcal{A}} = \mathcal{K}_2\mathcal{N}^2 - \mathcal{K}_3\mathcal{A}^2 = 0\,,
\end{equation}
which fully determines \(\mathcal{K}_1\) and \(\mathcal{K}_4\) in terms of the parameter functions \(\mathcal{N}\) and \(\mathcal{A}\) of the metric, giving
\begin{equation}\label{eq:metcompcond}
\mathcal{K}_1 = \frac{\dot{\mathcal{N}}}{\mathcal{N}}\,, \quad
\mathcal{K}_4 = \frac{\dot{\mathcal{A}}}{\mathcal{A}}\,,
\end{equation}
and establishes that \(\mathcal{K}_2\) and \(\mathcal{K}_3\) must be proportional to each other, with a positive factor \(\mathcal{A}^2/\mathcal{N}^2\) imposed by the condition that the metric is of Lorentzian signature. In the remainder of this section we will therefore restrict our attention to connections for which these conditions are satisfied.

\subsubsection{Flatness condition}\label{sssec:maflat}
To further restrict the connections under consideration, we now impose the flatness condition~\eqref{eq:affcurv}. Observe that the curvature of the general cosmologically symmetric connection is given by
\begin{equation}
\begin{split}
R_{\mu\nu\rho\sigma} &= 2\frac{\mathcal{K}_3(\mathcal{K}_4 - \mathcal{K}_1) + \dot{\mathcal{K}}_3}{\mathcal{N}^2}n_{\nu}n_{[\rho}h_{\sigma]\mu}
- 2\frac{\mathcal{K}_3\mathcal{K}_5}{\mathcal{N}\mathcal{A}}n_{\nu}\varepsilon_{\mu\rho\sigma}\\
&\phantom{=}+ 2\frac{\mathcal{K}_2(\mathcal{K}_4 - \mathcal{K}_1) - \dot{\mathcal{K}}_2}{\mathcal{A}^2}n_{\mu}n_{[\rho}h_{\sigma]\nu}
- 2\frac{\dot{\mathcal{K}}_5}{\mathcal{N}\mathcal{A}}\varepsilon_{\mu\nu[\rho}n_{\sigma]}\\
&\phantom{=}+ 2\frac{\mathcal{K}_2\mathcal{K}_5\mathcal{N}}{\mathcal{A}^3}n_{\mu}\varepsilon_{\nu\rho\sigma}
+ 2\frac{u^2 + \mathcal{K}_2\mathcal{K}_3 - \mathcal{K}_5^2}{\mathcal{A}^2}h_{\mu[\rho}h_{\sigma]\nu}\,,
\end{split}
\end{equation}
and so the connection is flat if and only if
\begin{multline}\label{eq:flatcond}
0 = \dot{\mathcal{K}}_5 = \mathcal{K}_2\mathcal{K}_5 = \mathcal{K}_3\mathcal{K}_5 = u^2 + \mathcal{K}_2\mathcal{K}_3 - \mathcal{K}_5^2\\
= \mathcal{K}_3(\mathcal{K}_4 - \mathcal{K}_1) + \dot{\mathcal{K}}_3 = \mathcal{K}_2(\mathcal{K}_4 - \mathcal{K}_1) - \dot{\mathcal{K}}_2\,.
\end{multline}
To determine the flat, metric-compatible connections it is helpful to first eliminate \(\mathcal{K}_1\) and \(\mathcal{K}_4\) using the metricity condition~\eqref{eq:metcompcond}, and then to solve the remaining equations. This is most easily done by distinguishing two cases:
\begin{enumerate}
\item
\(u = 0\): In this case we have the condition \(\mathcal{K}_2\mathcal{K}_3 = \mathcal{K}_5^2\), so either both sides are vanishing or non-vanishing. However, from \(\mathcal{K}_2\mathcal{K}_5 = \mathcal{K}_3\mathcal{K}_5 = 0\) follows that \(\mathcal{K}_5 = 0\) or \(\mathcal{K}_2 = \mathcal{K}_3 = 0\). Hence, the only option is \(\mathcal{K}_5 = \mathcal{K}_2\mathcal{K}_3 = 0\). Therefore, at least one of \(\mathcal{K}_2\) or \(\mathcal{K}_3\) must vanish. From metric compatibility follows that if one of them vanishes, so does the other, and so the only possibility is \(\mathcal{K}_2 = \mathcal{K}_3 = 0\).
\item
\(u \neq 0\):
We can distinguish two cases:
\begin{enumerate}
\item
\(\mathcal{K}_5 \neq 0\): From \(\mathcal{K}_2\mathcal{K}_5 = \mathcal{K}_3\mathcal{K}_5 = 0\) follows \(\mathcal{K}_2 = \mathcal{K}_3 = 0\). Hence, \(\mathcal{K}_5 = \pm u\), and the remaining equations are satisfied. The connection is real if and only if \(u\) is real, and so \(u^2 > 0\).
\item
\(\mathcal{K}_5 = 0\): Now one has \(\mathcal{K}_2\mathcal{K}_3 = -u^2 \neq 0\). Together with metric compatibility this yields
\begin{equation}\label{eq:flatvecbranch}
\mathcal{K}_2 = \pm iu\frac{\mathcal{A}}{\mathcal{N}}\,, \quad
\mathcal{K}_3 = \pm iu\frac{\mathcal{N}}{\mathcal{A}}\,,
\end{equation}
where the same sign must be chosen for both terms. Note that this is real only if \(u\) is imaginary and hence \(u^2 < 0\).
\end{enumerate}
\end{enumerate}
The conditions of metricity and flatness thus completely determine the parameter functions \(\mathcal{K}_1, \ldots, \mathcal{K}_5\) in the affine connection, up to choosing one of two discrete branches in the case \(u \neq 0\), where only one of them is real, depending on the sign of \(u^2\). One further sees that the formulas determining the connection derived for the two branches with \(u \neq 0\) also hold in the case \(u = 0\), where they jointly reduce to \(\mathcal{K}_2 = \mathcal{K}_3 = \mathcal{K}_5 = 0\) for both branches, so that we do not need to treat this case separately, as it is already included as a limiting case. In summary, we find that the non-vanishing connection coefficients are given by
\begin{gather}
\Gamma^t{}_{tt} = \frac{\dot{\mathcal{N}}}{\mathcal{N}}\,, \quad
\Gamma^r{}_{rt} = \Gamma^{\vartheta}{}_{\vartheta t} = \Gamma^{\varphi}{}_{\varphi t} = \frac{\dot{\mathcal{A}}}{\mathcal{A}}\,, \quad
\Gamma^r{}_{rr} = \frac{u^2r}{\chi^2}\,,\nonumber\\
\Gamma^{\vartheta}{}_{r\vartheta} = \Gamma^{\vartheta}{}_{\vartheta r} = \Gamma^{\varphi}{}_{r\varphi} = \Gamma^{\varphi}{}_{\varphi r} = \frac{1}{r}\,, \quad
\Gamma^r{}_{\vartheta\vartheta} = -r\chi^2\,, \quad
\Gamma^r{}_{\varphi\varphi} = -r\chi^2\sin^2\vartheta\,,\nonumber\\
\Gamma^{\varphi}{}_{\vartheta\varphi} = \Gamma^{\varphi}{}_{\varphi\vartheta} = \cot\vartheta\,, \quad
\Gamma^{\vartheta}{}_{\varphi\varphi} = -\sin\vartheta\cos\vartheta\,,
\end{gather}
which are identical for both branches, and for \(u \neq 0\) in addition either the components
\begin{gather}
\Gamma^r{}_{\varphi\vartheta} = -\Gamma^r{}_{\vartheta\varphi} = \pm ur^2\chi\sin\vartheta\,,\nonumber\\
\Gamma^{\vartheta}{}_{r\varphi} = -\Gamma^{\vartheta}{}_{\varphi r} = \frac{\pm u\sin\vartheta}{\chi}\,, \quad
\Gamma^{\varphi}{}_{r\vartheta} = -\Gamma^{\varphi}{}_{\vartheta r} = \frac{\mp u}{\chi\sin\vartheta}\,,
\end{gather}
or the components
\begin{gather}
\Gamma^t{}_{rr} = \pm iu\frac{\mathcal{A}}{\mathcal{N}\chi^2}\,, \quad
\Gamma^t{}_{\vartheta\vartheta} = \pm iu\frac{\mathcal{A}}{\mathcal{N}}r^2\,, \quad
\Gamma^t{}_{\varphi\varphi} = \pm iu\frac{\mathcal{A}}{\mathcal{N}}r^2\sin^2\vartheta\,,\nonumber\\
\Gamma^r{}_{tr} = \Gamma^{\vartheta}{}_{t\vartheta} = \Gamma^{\varphi}{}_{t\varphi} = \pm iu\frac{\mathcal{N}}{\mathcal{A}}
\end{gather}
are non-vanishing, depending on the choice of the branch.

\subsubsection{Symmetric tetrads in the diagonal gauge}\label{sssec:madiagtet}
Having calculated the most general metric and flat, metric-compatible affine connection with cosmological symmetry, we can now determine the most general tetrad and spin connection representing this metric-affine geometry. It is the virtue of the symmetry definition outlined in section~\ref{ssec:stsym} that \emph{any} tetrad giving the metric~\eqref{eq:metcosmo}, together with the unique spin connection
\begin{equation}\label{eq:spinconn}
\omega^A{}_{B\rho} = e_B{}^{\nu}\left(\theta^A{}_{\mu}\Gamma^{\mu}{}_{\nu\rho} - \partial_{\rho}\theta^A{}_{\nu}\right)
\end{equation}
which satisfies the relation~\eqref{eq:affconn} for the chosen tetrad and the affine connection~\eqref{eq:affcosmo}, will satisfy the symmetry conditions for the cosmological symmetry. Hence, to determine the most general combination of a tetrad and corresponding spin connection, it is sufficient to choose one particular tetrad yielding the cosmologically symmetric metric, determine its corresponding spin connection, and then the most general tetrad is obtained by applying all possible Lorentz transformations to this pair. The easiest possible choice for this initial tetrad is the diagonal tetrad~\eqref{eq:tetdiag}. Inserting this tetrad together with the cosmologically symmetric connection~\eqref{eq:affcosmo} displayed in section~\ref{sssec:macosmosym} into the relation~\eqref{eq:spinconn} yields the non-vanishing components of the general, cosmologically symmetric spin connection
\begin{subequations}\label{eq:spconcos}
\begin{gather}
\omega'^0{}_{0t} = \mathcal{K}_1 - \frac{\dot{\mathcal{N}}}{\mathcal{N}}\,, \quad
\omega'^1{}_{1t} = \omega'^2{}_{2t} = \omega'^3{}_{3t} = \mathcal{K}_4 - \frac{\dot{\mathcal{A}}}{\mathcal{A}}\,,\label{eq:spccosdiag}\\
\omega'^0{}_{1r} = \frac{\mathcal{N}\mathcal{K}_2}{\mathcal{A}\chi}\,, \quad
\omega'^0{}_{2\vartheta} = \frac{\mathcal{N}\mathcal{K}_2r}{\mathcal{A}}\,, \quad
\omega'^0{}_{3\varphi} = \frac{\mathcal{N}\mathcal{K}_2r\sin\vartheta}{\mathcal{A}}\,,\label{eq:spcos0i}\\
\omega'^1{}_{0r} = \frac{\mathcal{A}\mathcal{K}_3}{\mathcal{N}\chi}\,, \quad
\omega'^2{}_{0\vartheta} = \frac{\mathcal{A}\mathcal{K}_3r}{\mathcal{N}}\,, \quad
\omega'^3{}_{0\varphi} = \frac{\mathcal{A}\mathcal{K}_3r\sin\vartheta}{\mathcal{N}}\,,\label{eq:spcosi0}\\
\omega'^2{}_{1\varphi} = -\omega'^1{}_{2\varphi} = \mathcal{K}_5r\sin\vartheta\,, \quad
\omega'^1{}_{3\vartheta} = -\omega'^3{}_{1\vartheta} = \mathcal{K}_5r\,, \quad
\omega'^3{}_{2r} = -\omega'^2{}_{3r} = \frac{\mathcal{K}_5}{\chi}\,,\label{eq:spcosps}\\
\omega'^2{}_{1\vartheta} = -\omega'^1{}_{2\vartheta} = \chi\,, \quad
\omega'^3{}_{1\varphi} = -\omega'^1{}_{3\varphi} = \chi\sin\vartheta\,, \quad
\omega'^3{}_{2\varphi} = -\omega'^2{}_{3\varphi} = \cos\vartheta\,.\label{eq:spcoscon}
\end{gather}
\end{subequations}
Imposing the metric compatibility conditions derived in section~\ref{sssec:mametriccomp}, we see that the diagonal components~\eqref{eq:spccosdiag} vanish identically, while the two triples~\eqref{eq:spcos0i} and~\eqref{eq:spcosi0} become equal, so that the spin connection becomes antisymmetric. Finally, from imposing the flatness conditions derived in section~\ref{sssec:maflat}, we find the following two branches:
\begin{enumerate}
\item
In the case \(\mathcal{K}_2 = \mathcal{K}_3 = 0\) and \(\mathcal{K}_5 = \pm u\), the components~\eqref{eq:spcos0i} and~\eqref{eq:spcosi0} vanish identically, while the only non-vanishing components~\eqref{eq:spcosps} and~\eqref{eq:spcoscon} yield the spin connection~\eqref{eq:spconsdf} which we obtained from the two-form representations.
\item
Choosing the branch \(\mathcal{K}_5 = 0\) together with the condition~\eqref{eq:flatvecbranch} instead, we find that the components~\eqref{eq:spcosps} vanish identically, while the remaining components reproduce the spin connection~\eqref{eq:spconvec} obtained from the vector representation.
\end{enumerate}
Note that also here we do not need to treat the case \(u = 0\) separately, as it appears as the joint limiting case, in which only the components~\eqref{eq:spcoscon} are non-vanishing. Hence, we arrive at the same result for the cosmologically symmetric spin connections in the diagonal gauge as from the representation theoretic approach in section~\ref{sssec:ardiagtet}.

\subsubsection{Symmetric tetrads in the Weitzenböck gauge}\label{sssec:mawbtet}
To complete our construction of the most general cosmologically symmetric geometries following the metric-affine approach, and fully recover the tetrads displayed in section~\ref{sssec:arwbtet} in the Weitzenböck gauge, we still need to find a local Lorentz transformation to transform each of the teleparallel branches of the cosmologically symmetric spin connection~\eqref{eq:spconcos} to a vanishing one. However, this task is trivially solved by realizing that the spin connections we obtained in section~\ref{sssec:madiagtet} are the same as in section~\ref{sssec:ardiagtet}, and so the Lorentz transformations~\eqref{eq:lortrvec} and~\eqref{eq:lortrsdf} satisfy this purpose, leading to the tetrads~\eqref{eq:tetwbvec} and~\eqref{eq:tetwbsdf} in the Weitzenböck gauge. Hence, we reproduce the result obtained in section~\ref{sssec:arwbtet}, which finally proves the full equivalence of the two approaches. Note that the Weitzenböck gauge can equivalently be obtained by applying a global Lorentz transformation in addition to the local Lorentz transformation~\eqref{eq:lortrvec} or~\eqref{eq:lortrsdf} for the chosen branch. The resulting tetrad in the Weitzenböck gauge then also differs from the corresponding tetrad~\eqref{eq:tetwbvec} or~\eqref{eq:tetwbsdf}. Also this is consistent with the approach in section~\ref{ssec:algebra}, since the homomorphisms \(\boldsymbol{\Lambda}\) we employed are unique only up to conjugation by a global Lorentz transformation.

\subsection{Irreducible torsion decomposition approach}\label{ssec:decomp}
We finally come to the third approach, which is based on an irreducible decomposition of the torsion tensor, which is the only tensorial quantity characterizing the teleparallel geometry besides the metric tensor. We briefly review this decomposition and introduce the notation we use in section~\ref{sssec:idtordec}. We then impose cosmological symmetry on the irreducible components in section~\ref{sssec:idcosmo}, showing that they are fully determined by a scalar and a pseudo-scalar function of time. In section~\ref{sssec:idaffconn}, we derive the corresponding affine connection, and impose its flatness in section~\ref{sssec:idflat}. This step will finally prove that we obtain the same connection as from the metric-affine approach shown in section~\ref{ssec:metaffine}, so that the same procedure can be used to calculate the symmetric tetrads and spin connections.

\subsubsection{Irreducible torsion decomposition}\label{sssec:idtordec}
The third and last approach we present here makes use of the fact that the torsion tensor, due to its antisymmetry in the last two indices, decomposes into three irreducible parts in the form~\cite{Hehl:1994ue}
\begin{equation}
T^{\mu}{}_{\nu\rho} = \mathfrak{V}^{\mu}{}_{\nu\rho} + \mathfrak{A}^{\mu}{}_{\nu\rho} + \mathfrak{T}^{\mu}{}_{\nu\rho}\,,
\end{equation}
where the three parts are uniquely determined by the conditions
\begin{equation}
\mathfrak{A}^{\nu}{}_{\nu\mu} = \mathfrak{T}^{\nu}{}_{\nu\mu} = 0\,, \quad
\mathfrak{V}_{[\mu\nu\rho]} = \mathfrak{T}_{[\mu\nu\rho]} = 0\,.
\end{equation}
They can explicitly be determined by defining
\begin{subequations}
\begin{align}
\mathfrak{v}_{\mu} &= T^{\nu}{}_{\nu\mu}\,, &
\mathfrak{V}^{\mu}{}_{\nu\rho} &= \frac{2}{3}\delta^{\mu}_{[\nu}\mathfrak{v}_{\rho]}\,,\\
\mathfrak{a}_{\mu} &= \frac{1}{6}\epsilon_{\mu\nu\rho\sigma}T^{\nu\rho\sigma}\,, &
\mathfrak{A}_{\mu\nu\rho} &= \epsilon_{\mu\nu\rho\sigma}\mathfrak{a}^{\sigma}\,,\\
\mathfrak{t}_{\mu\nu\rho} &= T_{(\mu\nu)\rho} + \frac{1}{3}\left(T^{\sigma}{}_{\sigma(\mu}g_{\nu)\rho} - T^{\sigma}{}_{\sigma\rho}g_{\mu\nu}\right)\,, &
\mathfrak{T}^{\mu}{}_{\nu\rho} &= \frac{4}{3}\mathfrak{t}^{\mu}{}_{[\nu\rho]}\,,
\end{align}
\end{subequations}
where \(\mathfrak{v}_{\mu}\) and \(\mathfrak{a}_{\mu}\) are a vector and pseudo-vector, each having four independent components, while the tensor \(\mathfrak{t}_{\mu\nu\rho}\) satisfies the conditions
\begin{equation}\label{eq;tensorcond}
\mathfrak{t}_{[\mu\nu]\rho} = \mathfrak{t}_{(\mu\nu\rho)} = 0\,, \quad
\mathfrak{t}^{\nu}{}_{\nu\mu} = \mathfrak{t}^{\nu}{}_{\mu\nu} = \mathfrak{t}_{\mu\nu}{}^{\nu} = 0\,,
\end{equation}
and therefore has 16 independent components. We will make use of this decomposition in order to determine the most general cosmologically symmetric torsion tensor, and in turn the most general cosmologically symmetric teleparallel geometry.

\subsubsection{Cosmologically symmetric torsion}\label{sssec:idcosmo}
In the next step we apply the condition of cosmological symmetry, i.e., homogeneity and isotropy, to the irreducible components of the torsion tensor displayed above, i.e., we demand that their Lie derivatives with respect to the cosmological symmetry generators~\eqref{eq:genrot} and~\eqref{eq:gentra} vanish. For the vector \(\mathfrak{v}_{\mu}\) and the pseudo-vector \(\mathfrak{a}_{\mu}\) this condition implies that they must be proportional to the time direction, represented by the unit normal vector \(n_{\mu}\), so that
\begin{subequations}\label{eq:torcosdec}
\begin{align}
\mathfrak{v}_{\mu} &= 3\mathcal{T}_1n_{\mu}\,, &
\mathfrak{V}_{\mu\nu\rho} &= 2\mathcal{T}_1h_{\mu[\nu}n_{\rho]}\,,\\
\mathfrak{a}_{\mu} &= -2\mathcal{T}_2n_{\mu}\,, &
\mathfrak{A}_{\mu\nu\rho} &= 2\mathcal{T}_2\varepsilon_{\mu\nu\rho}\,,
\end{align}
\end{subequations}
where \(\mathcal{T}_1\) and \(\mathcal{T}_2\) are functions of the cosmological time coordinate \(t\) only, and the constant numerical factors are chosen for later convenience~\cite{Iosifidis:2020gth}. For the tensor component \(\mathfrak{t}_{\mu\nu\rho}\), the cosmological symmetry together with the conditions~\eqref{eq;tensorcond} implies that this contribution to the torsion tensor must vanish identically. Hence, the most general cosmologically symmetric torsion tensor is given by
\begin{equation}\label{eq:torcosmo}
T_{\mu\nu\rho} = 2\mathcal{T}_1h_{\mu[\nu}n_{\rho]} + 2\mathcal{T}_2\varepsilon_{\mu\nu\rho}\,,
\end{equation}
where \(\mathcal{T}_1(t)\) is a scalar function, while \(\mathcal{T}_2(t)\) is a pseudo-scalar function with respect to spatial reflection.

\subsubsection{Contortion and connection coefficients}\label{sssec:idaffconn}
Following the mathematical preliminaries summarized in section~\ref{ssec:telegeo}, we can now uniquely determine the coefficients \(\Gamma^{\mu}{}_{\nu\rho}\) of the teleparallel affine connection from the cosmologically symmetric torsion~\eqref{eq:torcosmo}. For this purpose, we first make use of the relation~\eqref{eq:contor} to calculate the contortion, which reads
\begin{equation}\label{eq:contcosmo}
K_{\mu\nu\rho} = 2\mathcal{T}_1h_{\rho[\mu}n_{\nu]} - \mathcal{T}_2\varepsilon_{\mu\nu\rho}\,.
\end{equation}
Together with the Levi-Civita connection of the metric~\eqref{eq:metcosmo}, whose non-vanishing components are given by
\begin{gather}
\lc{\Gamma}^t{}_{tt} = \frac{\dot{\mathcal{N}}}{\mathcal{N}}\,, \quad
\lc{\Gamma}^r{}_{tr} = \Gamma^{\vartheta}{}_{t\vartheta} = \Gamma^{\varphi}{}_{t\varphi} = \Gamma^r{}_{rt} = \Gamma^{\vartheta}{}_{\vartheta t} = \Gamma^{\varphi}{}_{\varphi t} = \frac{\dot{\mathcal{A}}}{\mathcal{A}}\,,\nonumber\\
\lc{\Gamma}^t{}_{rr} = \frac{\mathcal{A}\dot{\mathcal{A}}}{\mathcal{N}^2\chi^2}\,, \quad
\lc{\Gamma}^t{}_{\vartheta\vartheta} = \frac{\mathcal{A}\dot{\mathcal{A}}}{\mathcal{N}^2}r^2\,, \quad
\lc{\Gamma}^t{}_{\varphi\varphi} = \frac{\mathcal{A}\dot{\mathcal{A}}}{\mathcal{N}^2}r^2\sin^2\vartheta\,, \quad
\lc{\Gamma}^r{}_{rr} = \frac{u^2r}{\chi^2}\,,\nonumber\\
\lc{\Gamma}^{\vartheta}{}_{r\vartheta} = \Gamma^{\vartheta}{}_{\vartheta r} = \Gamma^{\varphi}{}_{r\varphi} = \Gamma^{\varphi}{}_{\varphi r} = \frac{1}{r}\,, \quad
\lc{\Gamma}^{\varphi}{}_{\vartheta\varphi} = \Gamma^{\varphi}{}_{\varphi\vartheta} = \cot\vartheta\,,\nonumber\\
\lc{\Gamma}^{\vartheta}{}_{\varphi\varphi} = -\sin\vartheta\cos\vartheta\,, \quad
\lc{\Gamma}^r{}_{\vartheta\vartheta} = -r\chi^2\,, \quad
\lc{\Gamma}^r{}_{\varphi\varphi} = -r\chi^2\sin^2\vartheta\,,
\end{gather}
we find that the connection is of the cosmologically symmetric form~\eqref{eq:affcosmo}, as one would expect, and that the parameter functions are given by
\begin{equation}\label{eq:metcompcconn}
\mathcal{K}_1 = \frac{\dot{\mathcal{N}}}{\mathcal{N}}\,, \quad
\mathcal{K}_2 = \frac{\mathcal{A}\dot{\mathcal{A}}}{\mathcal{N}^2} - \frac{\mathcal{A}^2\mathcal{T}_1}{\mathcal{N}}\,, \quad
\mathcal{K}_3 = \frac{\dot{\mathcal{A}}}{\mathcal{A}} - \mathcal{N}\mathcal{T}_1\,, \quad
\mathcal{K}_4 = \frac{\dot{\mathcal{A}}}{\mathcal{A}}\,, \quad
\mathcal{K}_5 = \mathcal{A}\mathcal{T}_2\,.
\end{equation}
Note that, by construction, the connection is metric-compatible, and so it satisfies the metricity conditions~\eqref{eq:metcompcosmo}. Hence, the resulting geometry belongs to the class of Riemann-Cartan spacetimes. However, in this general form it still has also curvature, and is thus not yet teleparallel.

\subsubsection{Flatness condition}\label{sssec:idflat}
We finally need to restrict the class of metric-compatible connections with coefficients~\eqref{eq:metcompcconn} to those which satisfy the flatness conditions~\eqref{eq:flatcond}. The procedure is identical to that detailed in section~\ref{sssec:maflat}, and so we will only summarize the implications for the parameter functions \(\mathcal{T}_1\) and \(\mathcal{T}_2\), using the same distinction between different cases.
\begin{enumerate}
\item
\(u = 0\): In this case we have \(\mathcal{K}_2 = \mathcal{K}_3 = \mathcal{K}_5 = 0\). The parameter functions are therefore given by
\begin{equation}
\mathcal{T}_1 = \frac{\dot{\mathcal{A}}}{\mathcal{N}\mathcal{A}}\,, \quad
\mathcal{T}_2 = 0\,.
\end{equation}
\item
\(u \neq 0\):
We can distinguish two cases:
\begin{enumerate}
\item
\(\mathcal{K}_5 \neq 0\): In this case we have \(\mathcal{K}_2 = \mathcal{K}_3 = 0\) and \(\mathcal{K}_5 = \pm u\), so that the parameter functions are given by
\begin{equation}
\mathcal{T}_1 = \frac{\dot{\mathcal{A}}}{\mathcal{N}\mathcal{A}}\,, \quad
\mathcal{T}_2 = \pm\frac{u}{\mathcal{A}}\,.
\end{equation}
\item
\(\mathcal{K}_5 = 0\): From the condition~\eqref{eq:flatvecbranch} follow immediately the parameter functions
\begin{equation}
\mathcal{T}_1 = \frac{\dot{\mathcal{A}}}{\mathcal{N}\mathcal{A}} \pm \frac{iu}{\mathcal{A}}\,, \quad
\mathcal{T}_2 = 0\,.
\end{equation}
\end{enumerate}
\end{enumerate}
Inserting the obtained values of the parameter functions for the different branches into the relations~\eqref{eq:metcompcconn}, we obtain the same connections which we have already determined in section~\ref{sssec:maflat}. One can thus apply the same procedure as in sections~\ref{sssec:madiagtet} and~\ref{sssec:mawbtet} in order to determine the symmetric tetrads and spin connections. We will not repeat these steps here, and conclude this section by stating that the result we obtained shows once more the mutual consistency of the three approaches we presented here.

\section{Coordinate transformations}\label{sec:cotrans}
The family of cosmological symmetries detailed in section~\ref{ssec:cosmosym} is invariant under a number of coordinate transformations. Hence, the same invariance is also inherited by the cosmological teleparallel geometries we discuss. The aim of this section is to briefly review these transformations and to display their actions on the two branches of teleparallel geometries we presented. In particular, we discuss the change of the time coordinate in section~\ref{ssec:cttime}, the global spatial rescaling in section~\ref{ssec:ctradial} and discrete spatial reflections in section~\ref{ssec:ctreflect}. Finally, we discuss a coordinate change of a different nature, which we still subsume under this section, since it makes use of the same mathematics: in section~\ref{ssec:cthyper}, we transform our teleparallel geometries to hyperspherical coordinates, in order to demonstrate their relation to earlier results.

\subsection{Change of the time coordinate}\label{ssec:cttime}
We start by discussing how the teleparallel geometries presented in section~\ref{sec:construction}, which are parametrized by the cosmological symmetry parameter \(u\), the lapse function \(\mathcal{N}(t)\) and scale factor \(\mathcal{A}(t)\), are related to each other by a change of the time coordinate \(t \mapsto \tilde{t}(t)\). The aim of this section is mostly illustrative. It is obvious that the generating vector fields~\eqref{eq:genrot} and~\eqref{eq:gentra} retain their form under this family of coordinate transformation, which also manifests itself in the fact that they commute with any vector field \(X = X^t(t)\partial_t\) generating infinitesimal changes of the time coordinate. Writing the transformation of the tetrad components as
\begin{equation}\label{eq:tettrans}
\tilde{\theta}^A(\tilde{x}(x)) = \tilde{\theta}^A{}_{\mu}(\tilde{x}(x))\,\dd\tilde{x}^{\mu} = \theta^A{}_{\mu}(x)\,\dd x^{\mu} = \theta^A(x)\,,
\end{equation}
one immediately reads off from the tetrads~\eqref{eq:tetwbvec} and~\eqref{eq:tetwbsdf} in the Weitzenböck gauge the relation
\begin{equation}\label{eq:lapsetimetrans}
\tilde{\mathcal{N}}(\tilde{t}(t))\,\dd\tilde{t} = \mathcal{N}(t)\,\dd t \quad \Rightarrow \quad \tilde{\mathcal{N}}(\tilde{t}(t)) = \left(\frac{\partial\tilde{t}}{\partial t}\right)^{-1}\mathcal{N}(t)\,,
\end{equation}
showing that the lapse is the component of a covector, while the scale factor transforms as a scalar, \(\tilde{\mathcal{A}}(\tilde{t}(t)) = \mathcal{A}(t)\), i.e., it is unchanged up to a change of its dependence on the now different time coordinate. This if of course the well-known transformation behavior of the lapse and scale factor of a Robertson-Walker metric~\eqref{eq:metcosmo}, which therefore fully describes also the change of the teleparallel geometry under a change of the time coordinate.

We finally remark that this relation does not depend on the choice of the Lorentz gauge, i.e., it retains its form also in any other gauge besides the Weitzenböck gauge. This holds true in particular also for the diagonal gauge. Indeed, one finds that the same relation~\eqref{eq:lapsetimetrans} for the transformation of the lapse function is also obtained from the diagonal tetrad~\eqref{eq:tetdiag}. Note that in this case also the coordinate transformation of the non-trivial spin connection must be taken into account; however, for the two branches~\eqref{eq:spconvec} and~\eqref{eq:spconsdf} the spin connection is invariant under a change of the time coordinate.

\subsection{Constant rescaling of the radial coordinate}\label{ssec:ctradial}
In cosmology it is most common to describe the spatially non-flat case in coordinates which are chosen such that the curvature parameter satisfies \(|k| = |u^2| = 1\). For the teleparallel geometries we derived, we have not made use of this normalization, and kept \(u\) arbitrary instead, in order to show that this yields two continuous families of geometries which intersect at the common value \(u = 0\). It is worth mentioning that these different normalizations are related by a rescaling
\begin{equation}
x \mapsto \tilde{x}\,, \quad
(t, r, \vartheta, \varphi) \mapsto (\tilde{t}, \tilde{r}, \tilde{\vartheta}, \tilde{\varphi}) = \left(t, \frac{r}{c}, \vartheta, \varphi\right)\,.
\end{equation}
of the radial coordinate by a positive constant \(c\). It follows that the non-vanishing components of the Jacobian are given by
\begin{equation}
\frac{\partial\tilde{t}}{\partial t} = \frac{\partial\tilde{\vartheta}}{\partial\vartheta} = \frac{\partial\tilde{\varphi}}{\partial\varphi} = 1\,, \quad
\frac{\partial\tilde{r}}{\partial r} = \frac{1}{c}\,.
\end{equation}
Under this transformation the rotation generators~\eqref{eq:genrot} retain their coordinate expressions, while the translation generators transform as
\begin{equation}
\tilde{T}_i(\tilde{x}(x)) = \tilde{T}_i^{\mu}(\tilde{x}(x))\tilde{\partial}_{\mu} = cT_i^{\mu}(x)\partial_{\mu} = cT_i(x)\,,
\end{equation}
where
\begin{subequations}
\begin{align}
\tilde{T}_1 &= \tilde{\chi}\sin\vartheta\cos\varphi\partial_{\tilde{r}} + \frac{\tilde{\chi}}{\tilde{r}}\cos\vartheta\cos\varphi\partial_{\vartheta} - \frac{\tilde{\chi}\sin\varphi}{\tilde{r}\sin\vartheta}\partial_{\varphi}\,,\\
\tilde{T}_2 &= \tilde{\chi}\sin\vartheta\sin\varphi\partial_{\tilde{r}} + \frac{\tilde{\chi}}{\tilde{r}}\cos\vartheta\sin\varphi\partial_{\vartheta} + \frac{\tilde{\chi}\cos\varphi}{\tilde{r}\sin\vartheta}\partial_{\varphi}\,,\\
\tilde{T}_3 &= \tilde{\chi}\cos\vartheta\partial_{\tilde{r}} - \frac{\tilde{\chi}}{\tilde{r}}\sin\vartheta\partial_{\vartheta}
\end{align}
\end{subequations}
and
\begin{equation}
\tilde{\chi}(\tilde{r}) = \sqrt{1 - \tilde{u}^2\tilde{r}^2} = \sqrt{1 - u^2r^2} = \chi(r)\,, \quad
\tilde{u} = cu\,.
\end{equation}
Their commutators accordingly take the form
\begin{equation}
[\tilde{T}_i, \tilde{T}_j] = [cT_i, cT_j] = c^2u^2\epsilon_{ijk}R_k = \tilde{u}^2\epsilon_{ijk}R_k\,.
\end{equation}
Using this transformation rule for \(u\), and writing the transformation of the tetrad again as~\eqref{eq:tettrans}, we have
\begin{equation}
\tilde{\mathcal{N}}(t) = \mathcal{N}(t)\,, \quad
\tilde{\mathcal{A}}(t) = c\mathcal{A}(t)\,,
\end{equation}
since \(\tilde{\mathcal{A}}\tilde{r} = \mathcal{A}r\), so that now \(\mathcal{A}\) transforms as the component of a one-form, as expected. Again, this does not depend on the choice of the Lorentz gauge.

\subsection{Spatial reflection}\label{ssec:ctreflect}
In section~\ref{ssec:cosmosym}, we have defined cosmological symmetry as invariance of the geometry under the action of a connected Lie group, which is generated by six vector fields. We now extend this notion by also including reflections. This leads to the following non-connected symmetry groups:
\begin{enumerate}
\item
For \(u^2 > 0\), we extend the symmetry group from \(\mathrm{SO}(4)\) to \(\mathrm{O}(4)\).
\item
For \(u^2 = 0\), we similarly extend the symmetry group from \(\mathrm{ISO}(3)\) to \(\mathrm{IO}(3)\).
\item
For \(u^2 < 0\), the situation is slightly different, since we extend the symmetry group from \(\mathrm{SO}_0(1,3)\) to the orthochronous (but not spatially orientation preserving) Lorentz group \(\mathrm{SO}^{\uparrow}(1,3)\). This is due to the fact that in this case the orbits of the symmetry group, which are the constant time \(t\) hypersurfaces of the spacetime manifold, are diffeomorphic to the connected component of a hyperbolic space, and time reflections would map between two different connected components of this space.
\end{enumerate}
Due to the homogeneity established by the translation generators~\eqref{eq:gentra}, reflection symmetry can be implemented by considering reflections preserving an arbitrary spacetime point, and so without loss of generality we choose the coordinate origin \(r = 0\) as the most simple choice. A point reflection would then be obtained by the coordinate transformation \(\vartheta \mapsto \pi - \vartheta, \varphi \mapsto \varphi + \pi\). However, note that the latter is simply a rotation around the polar axis, which is already included in the connected part of the cosmological symmetry group. Hence, we will omit it here and restrict ourselves to the equatorial reflection
\begin{equation}
x \mapsto \tilde{x}\,, \quad
(t, r, \vartheta, \varphi) \mapsto (\tilde{t}, \tilde{r}, \tilde{\vartheta}, \tilde{\varphi}) = (t, r, \pi - \vartheta, \varphi)\,.
\end{equation}
We find that the non-vanishing components of the Jacobian are given by
\begin{equation}
\frac{\partial\tilde{t}}{\partial t} = \frac{\partial\tilde{r}}{\partial r} = \frac{\partial\tilde{\varphi}}{\partial\varphi} = 1\,, \quad
\frac{\partial\tilde{\vartheta}}{\partial\vartheta} = -1\,,
\end{equation}
so that we obtain the following transformation rules for any tensor fields and connection coefficients:
\begin{enumerate}
\item
If the number of coordinate indices \(\vartheta\) on a tensor field of connection coefficient is odd, a factor \(-1\) is incurred.
\item
Due to the coordinate change \(\vartheta \mapsto \pi - \vartheta\), all occurrences of \(\cos\vartheta\) are replaced by \(-\cos\tilde{\vartheta}\), while \(\sin\vartheta\) is retained as \(\sin\tilde{\vartheta}\). This also propagates to constructed triangular functions such as \(\tan\vartheta\) or \(\cot\vartheta\).
\end{enumerate}
We can now apply these transformations to the cosmologically symmetric teleparallel geometries we have constructed. This is most easily done for the tetrads in the Weitzenböck gauge which we derived in section~\ref{sssec:arwbtet}. We find the following results:
\begin{enumerate}
\item
The tetrads~\eqref{eq:tetwbvec} obtained from the vector representation are invariant, except for the last component \(\theta^3\), which incurs a factor \(-1\). This factor, however, can be absorbed into a (non-proper) global Lorentz transformation \(\Lambda = \mathrm{diag}(1,1,1,-1)\), and so the geometry is invariant under reflections.
\item
For the tetrads~\eqref{eq:tetwbsdf} obtained from the self-dual and anti-self-dual two-form representations, we find that all terms containing a factor \(u\) change their sign, except for the last component \(\theta^3\), which in addition incurs a global factor \(-1\) on all terms. While the latter can again be absorbed into a global Lorentz transformation as in the vector case, this is not possible for the former. Instead, a reflection changes a tetrad obtained from the self-dual two-form representation to an anti-self-dual one and vice versa, and these are inequivalent geometries.
\end{enumerate}
Note that we can see these findings also in other geometric objects we derived:
\begin{enumerate}
\item
In the diagonal gauge derived in section~\ref{sssec:ardiagtet}, we see that the tetrad~\eqref{eq:tetdiag} is invariant up to a global Lorentz transformation \(\Lambda = \mathrm{diag}(1,1,-1,1)\), changing the sign of the component \(\theta'^2\). Applying the same Lorentz transformation to the non-vanishing spin connection changes the sign of each component carrying an odd number of indices \(2\). For the spin connections~\eqref{eq:spconvec} this transformation exactly cancels all factors incurred from the reflection, hence confirming the invariance of the geometry. This is not the case for the spin connections~\eqref{eq:spconsdf}, where all terms involving \(u\) incur a factor \(-1\), which cannot be absorbed into a Lorentz transformation, and relates the two inequivalent two-form representations.
\item
For the general cosmologically symmetric metric-affine geometry discussed in section~\ref{sssec:macosmosym}, we see that the FLRW metric~\eqref{eq:metcosmo} is invariant under reflections, while in the affine connection~\eqref{eq:affcosmo} all components involving the parameter function \(\mathcal{K}_5\) change their sign. Hence, this connection is invariant if and only if \(\mathcal{K}_5 = 0\). By comparison with the values of the parameter functions derived in section~\ref{sssec:maflat}, we find that this is the case for the branch corresponding to the vector representation, while for the two-form representations we have \(\mathcal{K}_5 = \pm u\).
\item
Following the irreducible decomposition of the torsion tensor displayed in section~\ref{sssec:idtordec}, one easily checks that the components of the axial part \(\mathfrak{A}^{\mu}{}_{\nu\rho}\) of the cosmologically symmetric torsion change their sign under a spatial reflection, while the remaining vector and tensor parts are invariant. This can be seen most easily from the form~\eqref{eq:torcosdec}, which shows that the vector part \(\mathfrak{V}_{\mu\nu\rho}\) is non-vanishing only if \(h_{\mu\nu}\) carries either two or zero indices \(\vartheta\), while \(\mathfrak{V}_{\mu\nu\rho}\) is non-vanishing only if \(\varepsilon_{\mu\nu\rho}\) carries exactly one index \(\vartheta\). Hence, the cosmologically symmetric torsion~\eqref{eq:torcosmo} is invariant under spatial reflections if and only if \(\mathcal{T}_2 = 0\). Comparing with the relations~\eqref{eq:metcompcconn}, we see that this is the case if and only if \(\mathcal{K}_5 = 0\), which agrees with our previous findings.
\end{enumerate}
Note that in teleparallel gravity theories which are symmetric under parity transformations, one would expect that spatial reflections have no effect on the resulting dynamics. However, it is also possible to construct teleparallel gravity theories which include parity-violating terms in their action, and for these one would expect different dynamics to arise from the use of the presented inequivalent tetrads.

\subsection{Solution in hyperspherical coordinates}\label{ssec:cthyper}
Though most of the literature on teleparallel cosmology makes use of either spherical coordinates as we have used so far, or Cartesian coordinates in case of a spatially flat geometry, there are also works which make use of hyperspherical coordinates instead. Using our conventions and notation these can most easily be defined as
\begin{equation}
r = S_u(\psi) = \psi\sinc(u\psi) = \sum_{n = 0}^{\infty}\frac{(-1)^nu^{2n}\psi^{2n+1}}{(2n + 1)!}\,,
\end{equation}
where
\begin{equation}
\sinc x = \begin{cases}
\frac{\sin x}{x} & x \neq 0\,,\\
1 & x = 0\,.
\end{cases}
\end{equation}
Using their convenient property
\begin{equation}
\frac{\dd r}{\chi} = \dd\psi\,, \quad
\chi\partial_r = \partial_{\psi}\,, \quad
\chi = C_u(\psi) = S_u'(\psi)\,,
\end{equation}
the diagonal tetrad~\eqref{eq:tetdiag} becomes
\begin{equation}
\theta'^0 = \mathcal{N}\dd t\,, \quad
\theta'^1 = \mathcal{A}\dd\psi\,, \quad
\theta'^2 = \mathcal{A}S_u(r)\dd\vartheta\,, \quad
\theta'^3 = \mathcal{A}S_u(r)\sin\vartheta\dd\varphi\,,
\end{equation}
while in the Weitzenböck gauge we have the vector tetrad~\eqref{eq:tetwbvec} represented as
\begin{subequations}
\begin{align}
\theta^0 &= \mathcal{N}C\dd t \pm iu\mathcal{A}S\dd\psi\,,\\
\theta^1 &= \mathcal{A}\left[\sin\vartheta\cos\varphi\left(C\dd\psi \pm iu\frac{\mathcal{N}}{\mathcal{A}}S\dd t\right) + S\cos\vartheta\cos\varphi\dd\vartheta - S\sin\vartheta\sin\varphi\dd\varphi\right]\,,\\
\theta^2 &= \mathcal{A}\left[\sin\vartheta\sin\varphi\left(C\dd\psi \pm iu\frac{\mathcal{N}}{\mathcal{A}}S\dd t\right) + S\cos\vartheta\sin\varphi\dd\vartheta + S\sin\vartheta\cos\varphi\dd\varphi\right]\,,\\
\theta^3 &= \mathcal{A}\left[\cos\vartheta\left(C\dd\psi \pm iu\frac{\mathcal{N}}{\mathcal{A}}S\dd t\right) - S\sin\vartheta\dd\vartheta\right]\,,
\end{align}
\end{subequations}
as well as the two-form case~\eqref{eq:tetwbsdf} given by
\begin{subequations}
\begin{align}
\theta^0 &= \mathcal{N}\dd t\,,\\
\theta^1 &= \mathcal{A}\bigg[\sin\vartheta\cos\varphi\dd\psi + S(C\cos\vartheta\cos\varphi \pm uS\sin\varphi)\dd\vartheta\nonumber\\
&\phantom{= \mathcal{A}\bigg[} \quad - S\sin\vartheta(C\sin\varphi \mp uS\cos\vartheta\cos\varphi)\dd\varphi\bigg]\,,\\
\theta^2 &= \mathcal{A}\bigg[\sin\vartheta\sin\varphi\dd\psi + S(C\cos\vartheta\sin\varphi \mp uS\cos\varphi)\dd\vartheta\nonumber\\
&\phantom{= \mathcal{A}\bigg[} \quad + S\sin\vartheta(C\cos\varphi \pm uS\cos\vartheta\sin\varphi)\dd\varphi\bigg]\,,\\
\theta^3 &= \mathcal{A}\left[\cos\vartheta\dd\psi - CS\sin\vartheta\dd\vartheta \mp uS^2\sin^2\vartheta\dd\varphi\right]\,,
\end{align}
\end{subequations}
using the abbreviations \(S = S_u(\psi)\) and \(C = C_u(\psi)\). Among the latter we indeed find the solution derived in~\cite{Capozziello:2018hly}.

\section{Cosmological dynamics}\label{sec:dynamics}
In order to demonstrate the use of the cosmological teleparallel geometries we derived on the previous sections, we now study the cosmological dynamics which arises from inserting the different tetrads and spin connections into the field equations of different teleparallel gravity theories. We start by discussing the general form of the cosmological field equations in the covariant formulation of teleparallel gravity in section~\ref{ssec:dyngeneral}. We then apply these considerations to two general classes of teleparallel gravity theories in section~\ref{ssec:dynpartic} and derive their cosmological field equations.

\subsection{General considerations}\label{ssec:dyngeneral}
Following the covariant approach to teleparallel gravity~\cite{Krssak:2018ywd}, the dynamical fields mediating the gravitational interaction are the tetrad \(\theta^A{}_{\mu}\) and the spin connection~\(\omega^A{}_{B\mu}\), while matter fields \(\Psi^I\) are minimally coupled to the metric induced by the tetrad and its Levi-Civita connection only. Assuming that there are no further, non-minimally coupled fields, the action thus takes the form
\begin{equation}
S[\theta, \omega, \Psi] = S_{\text{g}}[\theta, \omega] + S_{\text{m}}[\theta, \Psi]\,,
\end{equation}
where \(S_{\text{g}}\) and \(S_{\text{m}}\) denote the gravitational and matter parts, respectively. Variation of the former with respect to the tetrad yields the Euler-Lagrange expressions \(E_A{}^{\mu}\), while variation of the latter yields the energy-momentum tensor \(-\Theta_A{}^{\mu}\), so that, after lowering the second index and transforming the first index into a spacetime index using the tetrad, the field equations take the general form
\begin{equation}
E_{\mu\nu} = \Theta_{\mu\nu}\,.
\end{equation}
Further, from the assumption that matter couples minimally to the metric only follows that the matter action is locally Lorentz invariant, and hence the energy-momentum tensor is symmetric, \(\Theta_{[\mu\nu]} = 0\). For the gravitational part, local Lorentz invariance implies that variation of the field action with respect to the (flat) spin connection yields the antisymmetric part of the tetrad field equations, which are in general non-trivial. Hence, the total field equations split into a symmetric and antisymmetric part,
\begin{equation}
E_{(\mu\nu)} = \Theta_{\mu\nu}\,, \quad
E_{[\mu\nu]} = 0\,.
\end{equation}
The cosmological symmetry we study here significantly restricts the possible form of these two constituents. For the energy-momentum tensor, it mandates the perfect fluid form
\begin{equation}\label{eq:enmomcosmo}
\Theta_{\mu\nu} = (\rho + p)n_{\mu}n_{\nu} + pg_{\mu\nu} = \rho n_{\mu}n_{\nu} + ph_{\mu\nu}
\end{equation}
with matter density \(\rho\) and pressure \(p\). For the gravitational part, it similarly follows that the symmetric part must be of the form
\begin{equation}
E_{(\mu\nu)} = \mathfrak{N}n_{\mu}n_{\nu} + \mathfrak{H}h_{\mu\nu}\,,
\end{equation}
where \(\mathfrak{N}(t)\) and \(\mathfrak{H}(t)\) depend on the particular choice of the action, and can be expressed in terms of the parameter functions determining the cosmologically symmetric tetrad, while the antisymmetric part vanishes identically, \(E_{[\mu\nu]} \equiv 0\)~\cite{Hohmann:2019nat}. The cosmological field equations then simply read
\begin{equation}
\mathfrak{N} = \rho\,, \quad
\mathfrak{H} = p\,.
\end{equation}
It is instructive to calculate these two functions for a number of example theories and for the two tetrad branches we derived here. For this purpose it is most useful to write the torsion in the form~\eqref{eq:torcosmo} in terms of the scalar \(\mathcal{T}_1\) and the pseudo-scalar \(\mathcal{T}_2\), whose values for the different branches of cosmologically symmetric teleparallel geometries are given in section~\ref{sssec:idflat}. Further, we will set \(|u| = 1\) for the two spatially non-flat cases, and consider only the real solution branch for either sign of \(u^2\), in order to obtain a real action. Finally, realizing that
\begin{equation}
\frac{\dot{\mathcal{A}}}{\mathcal{N}\mathcal{A}} = \frac{\mathcal{L}_n\mathcal{A}}{\mathcal{A}} = H
\end{equation}
is the Hubble parameter, we will thus consider the cases
\begin{equation}\label{eq:torsflat}
\mathcal{T}_1 = H\,, \quad
\mathcal{T}_2 = 0
\end{equation}
for spatially flat FLRW spacetime,
\begin{equation}\label{eq:torspos}
\mathcal{T}_1 = H\,, \quad
\mathcal{T}_2 = \pm\frac{1}{\mathcal{A}}
\end{equation}
for positive spatial curvature \(k = 1\), as well as
\begin{equation}\label{eq:torsneg}
\mathcal{T}_1 = H \pm \frac{1}{\mathcal{A}}\,, \quad
\mathcal{T}_2 = 0
\end{equation}
for negative spatial curvature \(k = -1\). Also, for simplicity, we will make use of the time reparametrization freedom discussed in section~\ref{ssec:cttime}, and use the common cosmological time coordinate defined by \(\mathcal{N} \equiv 1\). Finally, note that most literature on teleparallel cosmology is based on the sign convention \((+,-,-,-)\) for the metric signature, in contrast to the convention \((-,+,+,+)\) we use here, and hence the obtained cosmological field equations differ by signs in several places.

\subsection{Particular theories}\label{ssec:dynpartic}
We now apply the general considerations on cosmological field equations displayed in the previous section to two commonly studied classes of teleparallel gravity. One of the most well-known classes, which is applied in particular to cosmology, is known as \(f(T)\) gravity, and will be discussed in section~\ref{sssec:dynft}. Another class, which is interesting in particular because of its different coupling of vector and axial torsion components, is known as new general relativity, and will be shown in section~\ref{sssec:dynngr}.

\subsubsection{$f(T)$ gravity}\label{sssec:dynft}
The first class of theories we consider is the so-called \(f(T)\) class of gravity theories~\cite{Bengochea:2008gz,Linder:2010py}. Using our choice of units for the gravitational constant, its action can be written in the form
\begin{equation}\label{eq:ftaction}
S_g = \frac{1}{2\kappa^2}\int\dd^4x\,\theta\,f(T)\,,
\end{equation}
where \(f\) is a free function of the torsion scalar \(T\), and the latter is defined as
\begin{equation}\label{eq:torsscal}
T = \frac{1}{2}T^{\rho}{}_{\mu\nu}S_{\rho}{}^{\mu\nu}\,,
\end{equation}
in terms of the superpotential
\begin{equation}\label{eq:suppot}
S_{\rho}{}^{\mu\nu} = K^{\mu\nu}{}_{\rho} - \delta_{\rho}^{\mu}T_{\sigma}{}^{\sigma\nu} + \delta_{\rho}^{\nu}T_{\sigma}{}^{\sigma\mu}\,.
\end{equation}
By variation with respect to the tetrad as discussed in the previous section, one obtains the gravitational part of the field equations
\begin{equation}\label{eq:ftfeq}
\kappa^2E_{\mu\nu} = \frac{1}{2}fg_{\mu\nu} + \lc{\nabla}_{\rho}\left(f_TS_{\nu\mu}{}^{\rho}\right) + f_TS^{\rho\sigma}{}_{\mu}\left(K_{\rho\nu\sigma} - T_{\rho\sigma\nu}\right)\,.
\end{equation}
In order to derive the cosmological field equations, we now make use of the relations~\eqref{eq:torcosmo} and~\eqref{eq:contcosmo} for the cosmologically symmetric torsion tensor and its corresponding contortion tensor. Inserting these into the definition~\eqref{eq:suppot} of the superpotential one finds
\begin{equation}
S_{\rho}{}^{\mu\nu} = 4\mathcal{T}_1n^{[\mu}h^{\nu]}_{\rho} - \mathcal{T}_2\varepsilon^{\mu\nu}{}_{\rho}\,,
\end{equation}
while for the torsion scalar~\eqref{eq:torsscal} one obtains the form
\begin{equation}
T = 6(\mathcal{T}_1^2 - \mathcal{T}_2^2)\,.
\end{equation}
With these expressions at hand, one can now evaluate the right hand side of the field equations~\eqref{eq:ftfeq}. For this purpose, it is helpful to note the expressions
\begin{subequations}
\begin{align}
S^{\rho\sigma}{}_{\mu}\left(K_{\rho\nu\sigma} - T_{\rho\sigma\nu}\right) &= -\frac{1}{3}Th_{\mu\nu} = 2(\mathcal{T}_2^2 - \mathcal{T}_1^2)h_{\mu\nu}\,,\\
\lc{\nabla}_{\rho}S_{\nu\mu}{}^{\rho} &= 6\frac{\mathcal{T}_1\mathcal{L}_n\mathcal{A}}{\mathcal{A}}n_{\mu}n_{\nu} - \frac{2\mathcal{L}_n(\mathcal{T}_1\mathcal{A}^2)}{\mathcal{A}^2}h_{\mu\nu}\,,\\
S_{\nu\mu}{}^{\rho}\lc{\nabla}_{\rho}T &= 24\mathcal{T}_1\left(\mathcal{T}_2\mathcal{L}_n\mathcal{T}_2 - \mathcal{T}_1\mathcal{L}_n\mathcal{T}_1\right)h_{\mu\nu}\,.
\end{align}
\end{subequations}
Together with the energy-momentum tensor~\eqref{eq:enmomcosmo}, one thus finds the general cosmological field equations
\begin{subequations}
\begin{align}
\kappa^2\rho &= -\frac{1}{2}f + 6f_T\frac{\mathcal{T}_1\mathcal{L}_n\mathcal{A}}{\mathcal{A}}\,,\\
\kappa^2p &= \frac{1}{2}f + 2f_T(\mathcal{T}_2^2 - \mathcal{T}_1^2) - 2f_T\frac{\mathcal{L}_n(\mathcal{T}_1\mathcal{A}^2)}{\mathcal{A}^2} + 12f_{TT}\mathcal{T}_1\mathcal{L}_n(\mathcal{T}_2^2 - \mathcal{T}_1^2)\,.
\end{align}
\end{subequations}
In order to obtain the final result, we still need to replace \(\mathcal{T}_1\) and \(\mathcal{T}_2\) with their values for the different cosmologically symmetric tetrad branches. Using the expressions given in section~\ref{ssec:dyngeneral}, we obtain the following equations:
\begin{enumerate}
\item
For the spatially flat \(k = 0\) case~\eqref{eq:torsflat}:
\begin{subequations}
\begin{align}
\kappa^2\rho &= -\frac{1}{2}f + 6f_TH^2\,,\\
\kappa^2p &= \frac{1}{2}f - 2f_T(\dot{H} + 3H^2) - 24f_{TT}H^2\dot{H}\,.
\end{align}
\end{subequations}
These equations have been thoroughly studied in the literature; see~\cite{Cai:2015emx} for an extensive review and~\cite{Hohmann:2017jao} for an analysis using the method of dynamical systems.
\item
For the spatially positively curved \(k = 1\) case~\eqref{eq:torspos}:
\begin{subequations}
\begin{align}
\kappa^2\rho &= -\frac{1}{2}f + 6f_TH^2\,,\\
\kappa^2p &= \frac{1}{2}f - 2f_T\left(\dot{H} + 3H^2 - \frac{1}{\mathcal{A}^2}\right) - 24f_{TT}H^2\left(\dot{H} + \frac{1}{\mathcal{A}^2}\right)\,.
\end{align}
\end{subequations}
These equations have also been found in~\cite{Capozziello:2018hly}, and applied also to the case \(k = -1\) by using a complex tetrad obtained from our solution branches by setting \(\mathcal{T}_2 = \pm i/\mathcal{A}\).
\item
For the spatially negatively curved \(k = -1\) case~\eqref{eq:torsneg}:
\begin{subequations}
\begin{align}
\kappa^2\rho &= -\frac{1}{2}f + 6f_TH\left(H \pm \frac{1}{\mathcal{A}}\right)\,,\\
\kappa^2p &= \frac{1}{2}f - 2f_T\left(\dot{H} + 3H^2 \pm 3\frac{H}{\mathcal{A}} + \frac{1}{\mathcal{A}^2}\right) - 24f_{TT}\left(H \pm \frac{1}{\mathcal{A}}\right)^2\left(\dot{H} \mp \frac{H}{\mathcal{A}}\right)\,.
\end{align}
\end{subequations}
These equations depend on the choice of the sign in the tetrad~\eqref{eq:tetwbvec} and thus lead to inequivalent cosmological dynamics. The lower choice of the sign has been derived in a more general class of scalar-torsion theories~\cite{Hohmann:2018rwf}, which generalizes the equations derived here.
\end{enumerate}
We see that the cosmological dynamics we obtain differ qualitatively between the different tetrad branches, and not only quantitatively by a term proportional to the spatial curvature parameter \(k\). In particular, we see that in the last case \(k = -1\) displayed above, even the choice of the sign in the tetrad~\eqref{eq:tetwbvec} leads to different cosmological dynamics, even though both tetrads represent the same FLRW metric. This clearly shows that the cosmological dynamics of \(f(T)\) gravity depends on degrees of freedom beyond the metric ones. It is thus even more remarkable, though not surprising, that in the general relativity limit \(f = T\), which implies \(f_T = 1\) and \(f_{TT} = 0\), all displayed cosmological field equations reduce to the Friedmann equations
\begin{equation}\label{eq:friedmann}
\kappa^2\rho = 3\left(H^2 + \frac{k}{\mathcal{A}^2}\right)\,, \quad
\kappa^2p = -2\dot{H} - 3H^2 - \frac{k}{\mathcal{A}^2}\,,
\end{equation}
where the different branches are distinguished only by the appearance of the spatial curvature parameter \(k\). This follows from the fact that in this case all tetrad degrees of freedom beyond the metric ones, and equivalently the spin connection, cancel from the field equations~\eqref{eq:ftfeq}, so that the latter reduce to Einstein's field equations, which depend only on the metric.

\subsubsection{New general relativity}\label{sssec:dynngr}
The second class of theories we consider is given by new general relativity~\cite{Hayashi:1979qx}. Its action can most easily be written by making use of the irreducible torsion decomposition outlined in section~\ref{sssec:idtordec} and reads
\begin{equation}\label{eq:ngraction}
S_g = \frac{1}{2\kappa^2}\int\dd^4x\,\theta\left(c_a\mathfrak{a}_{\mu}\mathfrak{a}^{\mu} + c_t\mathfrak{t}_{\mu\nu\rho}\mathfrak{t}^{\mu\nu\rho} + c_v\mathfrak{v}_{\mu}\mathfrak{v}^{\mu}\right)\,,
\end{equation}
with constants \(c_a, c_t, c_v\). Using the same decomposition, the gravity part of the field equations read
\begin{multline}
\kappa^2E_{\mu\nu} = c_a\left(\frac{1}{2}\mathfrak{a}^{\rho}\mathfrak{a}_{(\rho}g_{\mu\nu)} - \frac{4}{9}\epsilon_{\nu\alpha\beta\gamma}\mathfrak{a}^{\alpha}\mathfrak{t}_{\mu}{}^{\beta\gamma} - \frac{2}{9}\epsilon_{\mu\nu\rho\sigma}\mathfrak{a}^{\rho}\mathfrak{v}^{\sigma} + \frac{1}{3}\epsilon_{\mu\nu\rho\sigma}\lc{\nabla}^{\rho}\mathfrak{a}^{\sigma}\right)\\
+ c_t\left(\frac{2}{3}\mathfrak{t}_{\alpha[\beta\gamma]}\mathfrak{t}^{\alpha\beta\gamma}g_{\mu\nu} - \frac{4}{3}\mathfrak{t}_{\mu[\rho\sigma]}\mathfrak{t}_{\nu}{}^{\rho\sigma} + 2\lc{\nabla}^{\rho}\mathfrak{t}_{\mu[\nu\rho]} - \frac{2}{3}\mathfrak{t}_{\nu[\mu\rho]}\mathfrak{v}^{\rho} + \frac{1}{2}\epsilon_{\mu\alpha\beta\gamma}\mathfrak{a}^{\alpha}\mathfrak{t}_{\nu}{}^{\beta\gamma}\right)\\
+ c_v\left(\frac{1}{2}\mathfrak{v}^{\rho}\mathfrak{v}_{(\rho}g_{\mu\nu)} + \frac{4}{3}\mathfrak{t}_{\mu[\rho\nu]}\mathfrak{v}^{\rho} + 2g_{\mu[\nu}\lc{\nabla}^{\rho}\mathfrak{v}_{\rho]} - \frac{1}{2}\epsilon_{\mu\nu\rho\sigma}\mathfrak{a}^{\rho}\mathfrak{v}^{\sigma}\right)\,.
\end{multline}
It follows immediately from these expressions that the tensorial part governed by \(c_t\) does not contribute to the cosmological dynamics, where \(\mathfrak{t}_{\mu\nu\rho} = 0\), while the axial part proportional to \(c_a\) contributes only in the case of non-vanishing axial torsion. Expressed by the cosmologically symmetric metric~\eqref{eq:metcosmo} and torsion~\eqref{eq:torcosmo}, the only non-vanishing terms are given by
\begin{subequations}
\begin{align}
\mathfrak{a}^{\rho}\mathfrak{a}_{(\rho}g_{\mu\nu)} &= 4\mathcal{T}_2^2\left(n_{\mu}n_{\nu} - \frac{1}{3}h_{\mu\nu}\right)\,,\\
\mathfrak{v}^{\rho}\mathfrak{v}_{(\rho}g_{\mu\nu)} &= 9\mathcal{T}_1^2\left(n_{\mu}n_{\nu} - \frac{1}{3}h_{\mu\nu}\right)\,,\\
g_{\mu[\nu}\lc{\nabla}^{\rho}\mathfrak{v}_{\rho]} &= -\frac{9\mathcal{T}_1\mathcal{L}_n\mathcal{A}}{2\mathcal{A}}n_{\mu}n_{\nu} + \frac{3\mathcal{L}_n(\mathcal{T}_1\mathcal{A}^2)}{2\mathcal{A}^2}h_{\mu\nu}\,.
\end{align}
\end{subequations}
Hence, the cosmological equations of motion are given by
\begin{subequations}
\begin{align}
\kappa^2\rho &= -c_v\frac{9\mathcal{T}_1\mathcal{L}_n\mathcal{A}}{\mathcal{A}} + \frac{9}{2}c_v\mathcal{T}_1^2 + 2c_a\mathcal{T}_2^2\,,\\
\kappa^2 p &= c_v\frac{3\mathcal{L}_n(\mathcal{T}_1\mathcal{A}^2)}{\mathcal{A}^2} - \frac{3}{2}c_v\mathcal{T}_1^2 - \frac{2}{3}c_a\mathcal{T}_2^2\,.
\end{align}
\end{subequations}
For the different branches given in section~\ref{ssec:dyngeneral}, we thus find the following equations:
\begin{enumerate}
\item
For the spatially flat \(k = 0\) case~\eqref{eq:torsflat}:
\begin{equation}
\kappa^2\rho = -\frac{9}{2}c_vH^2\,, \quad
\kappa^2p = 3c_v\left(\dot{H} + \frac{3}{2}H^2\right)\,.
\end{equation}
\item
For the spatially positively curved \(k = 1\) case~\eqref{eq:torspos}:
\begin{equation}
\kappa^2\rho = -\frac{9}{2}c_vH^2 + \frac{2c_a}{\mathcal{A}^2}\,, \quad
\kappa^2p = 3c_v\left(\dot{H} + \frac{3}{2}H^2\right) - \frac{2c_a}{3\mathcal{A}^2}\,.
\end{equation}
\item
For the spatially negatively curved \(k = -1\) case~\eqref{eq:torsneg}:
\begin{equation}
\kappa^2\rho = -\frac{9}{2}c_v\left(H^2 - \frac{1}{\mathcal{A}^2}\right)\,, \quad
\kappa^2p = 3c_v\left(\dot{H} + \frac{3}{2}H^2 - \frac{1}{2\mathcal{A}^2}\right)\,.
\end{equation}
\end{enumerate}
It is remarkable that for the general relativity limit, which is given by the constants
\begin{equation}
c_v = -\frac{2}{3}\,, \quad
c_a = \frac{3}{2}\,, \quad
c_t = \frac{2}{3}\,,
\end{equation}
the equations for the different tetrad branches take the common form~\eqref{eq:friedmann}, and hence reduce to the well-known Friedmann equations, as already shown for \(f(T)\) gravity in the previous section. This once more supports the consistency of the presented approach.

\section{Conclusion}\label{sec:conclusion}
We have constructed the most general class of cosmologically symmetric, i.e., homogeneous and isotropic teleparallel geometries. For this construction we have applied three independent methods, and shown that they yield the same result, proving their mutual consistency. We have found that these geometries can be grouped into two branches, each of which is parametrized by a (real or imaginary) continuous parameter \(u\), whose square indicates the spatial curvature of the resulting FLRW metric, and that both branches intersect at \(u = 0\) to yield a unique spatially flat teleparallel geometry. For each branch of geometries we have provided explicit formulas for the tetrads in the Weitzenböck gauge, tetrad and spin connection in a diagonal gauge, metric-affine geometry, as well as torsion and contortion tensors.

Further, we have studied the behavior of the obtained geometries under coordinate transformations which are compatible with the cosmological symmetry and again provided explicit formulas. In particular, we have extended the notion of cosmological symmetry to also include spatial reflections and found that only the geometries belonging to one of the two branches are invariant under reflections, while for the other branch reflections relate inequivalent teleparallel geometries by changing the sign of their axial torsion. We speculated on the significance of this finding for theories whose action is not invariant under parity transformations.

In order to demonstrate a practical application of the teleparallel geometries we constructed, we used them in order to derive the cosmological dynamics of two classes of teleparallel gravity theories, namely \(f(T)\) gravity and new general relativity, and found that the different branches we obtained yield qualitatively different cosmological dynamics, unless we consider the general relativity limit of these theories. This behavior differs from other classes of gravity theories which are based on the curvature of the Levi-Civita connection of the metric tensor, where the effect of the spatial curvature of the FLRW metric is purely quantitative and parametrized by the constant spatial curvature parameter \(k = u^2\). Further studies of teleparallel gravity theories and their cosmological dynamics, generalizing the findings from~\cite{Hohmann:2017jao}, are required to show the physical significance of this result.

Another possible direction of future research beyond the cosmological background evolution is to study perturbations of the presented cosmologically symmetric geometries and their dynamics. This would extend and generalize previous works, where a spatially flat background tetrad has been assumed~\cite{Golovnev:2018wbh,Golovnev:2020aon}. The results could then be applied to numerous aspects of cosmology, such as the cosmic microwave background, the growth of density perturbations in structure formation or the propagation of gravitational waves on an expanding background geometry. Further, they would provide a first step towards developing the perturbation theory of more general teleparallel geometries, for example exhibiting only spherical symmetry~\cite{Hohmann:2019fvf}, which could in turn be applied to study the gravitational waves emitted by a nearly spherically symmetric source in terms of its quasi-normal modes.

\section*{Acknowledgments}
The author gratefully acknowledges the full support by the Estonian Research Council through the Personal Research Funding project PRG356, as well as the European Regional Development Fund through the Center of Excellence TK133 ``The Dark Side of the Universe''.

\end{document}